\documentclass[11pt]{article}
\usepackage{latexsym}
\usepackage{mathtools}
\usepackage[symbol]{footmisc}

\begin{document}
     \newcommand{\ncomm}{\newcommand}
     \ncomm{\ind}{\hspace*{.2in}}
     \ncomm{\noind}{\noindent}
     \ncomm{\hx}[1]{\hspace*{#1ex}}
     \ncomm{\tab}{\vspace*{-.15in}\begin{tabbing}}
     \ncomm{\untab}{\end{tabbing}\vspace*{-.1in}}
     \def\UL#1{$\underline{\smash{\hbox{{\it #1}}}}$}
     \ncomm{\al}{\forall}
     \ncomm{\ex}{\exists}
     \ncomm{\an}{\wedge}
     \ncomm{\ra}{\rightarrow}
     \ncomm{\la}{\leftarrow}
     \ncomm{\lc}{\langle}
     \ncomm{\rc}{\rangle}
     \ncomm{\Al}{$\forall$}
     \ncomm{\Ex}{$\exists$}
     \ncomm{\An}{$\wedge$}
     \ncomm{\Or}{$\vee$}
     \ncomm{\In}{$\in$}
     \ncomm{\Ra}{$\rightarrow$}
     \ncomm{\La}{$\leftarrow$}
     \ncomm{\RA}{$\Rightarrow$}
     \ncomm{\LA}{$\Leftarrow$}
     \ncomm{\Eq}{$\Leftrightarrow$}
     \ncomm{\Llra}{$\longleftrightarrow$}
     \ncomm{\Lc}{$\langle$}
     \ncomm{\Rc}{$\rangle\,$}
     \ncomm{\hs}[1]{\hspace*{#1in}}
     \ncomm{\vs}[1]{\vspace*{#1in}}
     \ncomm{\hm}[1]{\hspace*{#1em}}
     \ncomm{\vx}[1]{\vspace*{#1ex}}

     \ncomm{\NN}{I\hspace*{-.03in}N} 
     \ncomm{\Rr}{{\rm\it I\hspace*{-.06in}R}} 
     \ncomm{\RR}{{\rm\it I\hspace*{-.035in}R}} 
     \ncomm{\x}{\times}
     \ncomm{\X}{$\times$} 
     \ncomm{\pss}{\subset} 
     \ncomm{\Pss}{$\subset$}
     \ncomm{\sseq}{\subseteq}  
     \ncomm{\Sseq}{$\subseteq$}
     \ncomm{\px}{\sqsubseteq}  
     \ncomm{\Px}{$\sqsubseteq$}
     \ncomm{\ppx}{\sqsubset}   
     \ncomm{\Ppx}{$\sqsubset$}
     \ncomm{\xt}{\sqsupseteq}  
     \ncomm{\Xt}{$\sqsupseteq$}
     \ncomm{\pxt}{\sqsupset}   
     \ncomm{\Pxt}{$\sqsupset$}
     \ncomm{\emp}{\emptyset}
     \ncomm{\Emp}{$\emptyset$}
     \ncomm{\nil}{\Lambda} 
     \ncomm{\Nil}{$\Lambda$} 
     \ncomm{\oo}{\infty}
     \ncomm{\Oo}{$\infty$}
     \ncomm{\lm}{\lambda}
     \ncomm{\Lm}{$\lambda$}
     \ncomm{\Uni}{$\cup$}
     \ncomm{\Int}{$\cap$}
     \ncomm{\Mu}{$\mu$}
     \ncomm{\sig}{\sigma}
     \ncomm{\Sig}{$\sigma$}
     \ncomm{\PHI}{$\phi$}
     \ncomm{\codes}{\mathrlap{-}{>}} 
     \ncomm{\rcodes}{\mathrlap{-}{\gg}} 
     \ncomm{\limn}{\lim_{n}}
     \ncomm{\Limn}{$\lim_{n}$}
     \ncomm{\limsupn}{\limsup_{n}}
     \ncomm{\Limsupn}{$\limsup_{n}$}
     \ncomm{\liminfn}{\liminf_{n}}
     \ncomm{\Liminfn}{$\liminf_{n}$}
     \ncomm{\hh}{\hangindent=.25in} 

\vspace*{-.25in}
\begin{center}
{\LARGE {\bf Predictability and Randomness}}\\*[.05in]

  Lenhart K.\ Schubert\\
  University of Rochester
\end{center}

\noind
{\bf Abstract}. Algorithmic theories of randomness
can be related to theories of probabilistic sequence prediction
through the notion of a predictor, defined as a function
which supplies lower bounds on initial-segment probabilities
of infinite sequences. An infinite binary sequence $z$ is called
unpredictable iff its initial-segment ``redundancy" $n+\log p(z(n))$
remains sufficiently low relative to every effective predictor $p$.
A predictor which maximizes the initial-segment redundancy
of a sequence is called optimal for that sequence. It turns
out that a sequence is random iff it is unpredictable.
More generally, a sequence is random relative to an
arbitrary computable distribution iff the distribution
is itself an optimal predictor for the sequence. Here
``random" can be taken in the sense of Martin-L\"{o}f by 
using weak criteria of effectiveness, or in the sense of
Schnorr by using stronger criteria of effectiveness.
Under the weaker criteria of effectiveness it is possible
to construct a universal predictor which is optimal for 
all infinite sequences. This predictor assigns 
nonvanishing limit probabilities precisely to the 
recursive sequences. Under the stronger criteria of
effectiveness it is possible to establish a law of large
numbers for sequences random relative to a computable 
distribution, which may be useful as a criterion of
``rationality" for methods of probabilistic prediction.
A remarkable feature of effective predictors is the fact
that they are expressible in the special form first
proposed by Solomonoff. In this form sequence prediction
reduces to assigning high probabilities to initial segments
with short and/or numerous encodings. This fact provides
the link between theories of randomness and Solomonoff's
theory of prediction.

\newpage\noind
{\bf\large Preface}   

\noind
This article makes available an extended study of the theoretical 
relationship between predictability and randomness, for many years 
available only as a technical report in the Computing Science 
Department of the University of Alberta (TR77-2, September 1977, 
now no longer available). The typography in the TR was poor, as 
text formatting had not come of age yet. In essence the article 
shows that definitions of (non)randomness for infinite sequences in 
terms of computational nonrandomness tests, predictability, and 
``compressibility" through encodings as programs are equivalent.

I conducted the research leading to the report from 1975-1977, 
unaware except in the last months of the work that a 1970 article
in a Soviet journal had reported many of ``my" theorems, without 
proofs (Zvonkin \& Levin, 1970). I accordingly annotated those
theorems with (Leonid) Levin's name, before submitting the work 
for publication. The preempted theorems were limited to the 
semicomputable characterization of randomness, while my manuscript 
also covered the recursively computable characterization by C.P.\
Schnorr. However, while the manuscript was under review, a
treatment of the latter characterization appeared in an
article by that author.\footnote[1]%
  {Claus-Peter Schnorr, \& P. Fuchs, ``General Random Sequences and Learnable 
  Sequences,"  {\it J.\ Symb.\ Logic 42}(3), pp.\,329-340 (1977); also,\\
  \ind C.P.\ Schnorr, ``A survey of the theory of random sequences', in R.E.\ 
  Butts and J.\ Hintikka (eds.), {\it Basic Problems in Methodology and 
  Linguistics}, Dordrecht: D. Reidel, pp.\,193–210 (1977).}
The reviewer mentioned the possibility
of extending the article to a more complete survey, but I
found it difficult to contemplate presenting my hard-won results
as mostly a survey of results by others. Thus, the report 
languished as a TR, even though all the proofs were new, and 
a few unpublished propositions remained. Certain other researchers 
in this area later suggested to me that the missing proofs in 
the Zvonkin \& Levin survey should really have entitled me to
publication and co-discoverer status for at least the semicomputable 
results.  In any event, the existence of the arXiv system has made 
it possible to make the original version easily accessible.

The apparently new results in the submitted article were Theorems 
2-4 (on the extent to which semicomputable measures actually
allow probabilistic prediction), Th.\,6 (which seems to slightly
strengthen the previously known result that there is no recursive
universal distribution -- ``probabilities" assigned by ``predictors"
don't have to add up to 1), Th.\,7 (about a nearly optimal
{\it additive} ``predictor" -- a  nontrivial result, Th.\,11
(a kind of law of large numbers, which the reviewer said was 
implicit in some of Schnorr's published work, a comment that I 
did not succeed in confirming), and Th.\,13, with Corollaries 1 
\& 2 (though Levin had proved very closely related results).\footnote[7]%
{In 1988 I communicated these points, along with the TR, to Professor 
Levin, by then at Boston University.}

I include the symbol glossary that prefaced the TR, even though the
meanings of the symbols are mostly clear from the text.

\newpage\noind
{\bf\large Symbol Glossary}  
\tab
\bf{Sy}\=\bf{mbol} \hspace{1.8in} \=\bf{Meaning}\\*[.05in]
\> 0,1 \hspace{2.3in} \> unit strings; or numbers (clear from context)\\
\> $N$ \> \{0,1,2,...\}\\
\> \RR \> the nonnegative reals\\  
\> \RR$^{+}$ \> the positive reals\\ 
\> $R$ \> the real interval [0,1]\\
\> $B$ \> the numbers in $R$ with finite radix-2 representations\\
\> $Q$ \> the rational numbers in [0,1]\\
\> $X$ \> the 2-element alphabet \{0,1\}\\
\> $X^*$ \> the concatenation closure of $X$\\
\> $X^{\oo}$ \> the semi-infinite binary sequences\\ 
\> upper case Latin letters \> subsets of $X^*$ or $X^*\x \Rr$;\\
\> \hx{1} other than $B, N, R, Q, X$ \> \hx{1} or, procedure variables\\
\> dom \> domain\\
\> $f(S)$ \> $\{f(x) | x\in S\}$\\ 
\> $f^{-1}(S)$ \> $\{x | f(x) \in S\}$\\ 
\> $\log$ \> base 2 logarithm\\
\> $p, p', p'', p_{i}$ \> predictors or conditional predictors\\ 
\> $p^*(x)$ \> surplus probability of $x$ =$_{df}\;p(x)-p(x0)-p(x1)$\\
\> $p_{f}$ \> the Solomonoff predictor determined by a process $f$\\ 
\> pf \> prefix-free\\
\> re \> recursively enumerable\\
\> $rp(x)$ \> $|x| + \log p(x)$\\
\> $x$ and other lower case\\ \> \hx{1} letters near the end of\\
\>  \hx{1} the Latin alphabet \> binary sequence\\
\> $x(n)$ \> prefix of length $n$ of binary sequence $x$\\
\> $x^{-}$ \> finite binary sequence $x$  with last digit complemented\\ 
\> $x(n)^{-}$ \> $x(n)$ with $n$th digit complemented\\ 
\> $\delta$ \> an element of $\Rr^{+}$\\ 
\> $\lm$ \> Church's lambda operator\\ 
\> $\nil$ \> the null sequence\\  
\> $\mu$ \> a measure on subsets of $X^{\oo}$\\ 
\> $\sigma$ \> $\sigma S = \sum_{x\in S}{2^{-|x|}}$\\ 
\> $\phi$ \> partial recursive function from $N$ to $N$ or 
                                                   from $X^*\x N$ to $Q$\\ 
\> \Emp \> the empty set\\ 
\> \X \> Cartesian product\\ 
\> \Px \> is a prefix of\\  
\> \Ppx \> is a proper prefix of\\ 
\> \Xt \> is an extension of \\ 
\> \Pxt \> is a proper extension of\\ 
\> $(~)$ \> open interval; syntactic delimiters\\
\> [~] \> closed interval; assertion 
  delimiters (numerical value 1 or 0\\
  \>\>\hx{1}corresponding to true or false when used as arithmetic expression)\\
\> \Lc~\Rc \> ordered pair\\   
\> $\codes$ \> is an encoding of\\   
\> $\rcodes$ \> is a reduced encoding of  
\untab

\newpage
\subsection*{1. Introduction}
The decade beginning in 1963 saw the development of two types
of computational theories for infinite sequences.
Theories of the first type are concerned with the algorithmic
distinction between random and nonrandom sequences,
while theories of the second type are concerned with
prediction of infinite sequences or inductive discovery
of programs for them.

Relatively little attention has been paid 
to the connections between these two lines of development,
even though the existence of such connections has 
always been apparent. Indeed, von\,Mises'\,(1919)
original proposal for characterizing random sequences 
involved a notion similar to prediction, viz., {\it a priori}
selection of digits from an infinite binary sequence.
Von Mises' proposal, taken up by Wald~(1937) and
Church~(1940) among others, did not lead to a
satisfactory characterization of random sequences 
(see the critique of Ville, 1939).
The later work of Kolmogorov~(1965), 
Martin-L\"{o}f~(1966), Chaitin (1966), and Schnorr~(1971)
at last yielded several apparently successful
approaches to this problem.
However, none of these approaches turned explicitly
upon any proper notion of sequence prediction (although
the stakes wagered in Schnorr's ``gambling strategies"
could be viewed as implicit predictions).

On the other hand, the work on inductive inference
was not directly concerned with the definition of randomness.
Solomonoff~(1964) proposed
several classes of methods for predicting sequences
probabilistically, and Willis~(1970) showed that
one of these classes contains approximations to all recursive
sequential probability distributions.
Cover (1974), like Schnorr (1971), investigated sequential
gambling schemes. He explicitly related them to prediction
schemes and devised an interesting variant of one of
Solomonoff's universal prediction schemes. Subsequently 
Solomonoff (1976) proved convergence and other desirable
properties for his original universal predictor.

The studies most directly concerned with the relationship
between prediction and randomness are those of Chaitin
and Levin. Chaitin~(1975) defined randomness
in terms of Solomonoff-like probabilities, and has  
asserted (Chaitin, 1977) that his definition is
equivalent to that of Martin-L\"{o}f (1966).
At first sight Chaitin's probabilities seem unsuitable
for infinite sequence prediction: the probability assigned
to an initial segment may exceed the probabilities of
shorter initial segments. However, it seems clear in retrospect
that Chaitin's probabilities could have been used as a basis 
for the present study. The probability of a sequence as
discussed herein apparently corresponds to the sum of 
Chaitin's probabilities over all finite extensions of the
sequence.

The Soviet\footnote[2]{at the time} mathematician L.\,A.~Levin
made major contributions to the unification of the theories
of randomness and prediction in a series of papers sparked
by Levin's association with Kolmogorov (Zvonkin \& Levin, 1970,
and Levin, 1973, 1976).
Levin introduced the notion of a semicomputable measure,
which can be viewed intuitively as a method of probabilistic
sequence prediction. The theorems of Levin in Zvonkin and 
Levin (1970) show, in effect, that semicomputable measures are
expressible in the form proposed by Solomonoff (1964), although
Levin did not explicitly attach this interpretation to his 
results. Furthermore, he established analogous results for
computable measures (similar to the results obtained 
independently by Willis (1970)). Subsequently Levin (1973)
made the crucial connection between semicomputable measures
and randomness, stating a theorem to the effect that a sequence
is random in the sense of Martin-L\"{o}f iff it is irredundant
with respect to every semicomputable measure; indeed, he found
more generally that a sequence is Martin-L\"{o}f random
{\it relative} to any given computable distribution iff
its redundancy as measured by any semicomputable measure is
no greater than its redundancy relative to the given
distribution (apart from a constant). Later Levin (1976) further
generalized these results to sequences which are random
relative to arbitrary (not necessarily computable) measures,
and related them to information theory.

\renewcommand*{\thefootnote}{\arabic{footnote}}

The results of the present paper were obtained before the
author became aware of Levin's work. The central concern is with
sequence prediction in the sense of prior (``subjective")
probability assignments to initial segments of infinite sequences.
The objective is to relate this notion of prediction to definitions
of randomness due to Martin-L\"{o}f and Schnorr on the one hand and
to Solomonoff's ideas about prediction on the other. Several of
the main results of Secs.\,3 and 5 are contained in the cited papers
of Levin. The presentation of new proofs is justified in part
by the differences in approach (e.g., the construction of an
optimal predictor in Th.\,8 without reduction of predictors
to Solomonoff's form) and in part by the fact that Levin did not
publish proofs for all of his results (e.g., the connection
between predictability and Martin-L\"{o}f randomness, Th.\,5).
The present paper is more explicitly concerned with sequence
prediction than Levin's studies; the terminology and techniques
reflect this concern.

A topic not treated here is the extrapolation of
recursive sequences or inductive discovery of programs
for such sequences (e.g., Gold, 1967, and Blum \& Blum, 1973).
Although nonprobabilistic extrapolation of recursive sequences
can be viewed as a special case of probabilistic
prediction, the results herein are of too general a nature to 
shed any new light on this special case.

 Sec.\,2 introduces the formal notation and some
basic concepts. An incrementable predictor (cf.\ Levin's 
semicomputable measure) is defined as a lower bound on a sequential 
probability distribution which is approachable from below.
Thus the class of incrementable predictors contains all recursive
methods of prediction, as well as certain nonrecursive methods.
Alternative intuitive interpretations of predictors are
considered, and some computability properties of prediction
schemes based on incrementable predictors are examined.

In Sec.\,3 it is shown that any infinite binary sequence
$z$ is Martin-L\"{o}f random iff its initial-segment
redundancy $n+\log p(z(n))$ is bounded relative to every
incrementable predictor $p$. Actually this is established as a
corollary of the fact that $z$ is Martin-L\"{o}f random
{\it relative} to distribution $p$ iff $p$ is {\it optimal}
for $z$, where optimality means maximization of 
initial-segment redundancy.
An optimal {\it universal} predictor is then constructed,
i.e., one which maximizes the initial-segment
redundancy of every infinite binary sequence. This 
predictor assigns nonvanishing limit probabilities 
precisely to the recursive sequences. 

 In Sec.\,4 attention is restricted to
recursive predictors. It is shown that an infinite 
sequence is Schnorr-random iff its
initial-segment redundancy does not grow ``noticably"
(in a suitable sense) relative to any recursive predictor.
Again this is a corollary of a result about
Schnorr-randomness {\it relative} to a recursive
predictor, viz., that $z$ is Schnorr-random 
relative to $p$ iff $p$ is ``weakly optimal" for $z$. A related fact
is that a recursive predictor maximizes the initial-segment 
redundancy of a sequence only if the 
conditional probabilities it assigns to events
occurring in that sequence agree with the frequencies
of those events. For example, for about 70\% of the cases where
a 1-digit is predicted with 70\% conditional probability,
a 1-digit actually occurs. Thus an optimal predictor exhibits
a type of consistency which seems desirable in any sequential
inductive method.

 In Sec.\,5 it is shown that the class of incrementable
predictors coincides with one of Solomonoff's classes of predictive
methods based on program lengths. Also some variants of
Willis'\,1970) and Levin's (Zvonkin \& Levin, 1970) results about
the reduction of recursive predictors to Solomonoff's form
are presented.

\subsection*{2. Predictors}
 The following basic notation and terminology will be used.
$N$ is the set of natural numbers including 0,
\RR\ is the set of nonnegative real numbers, $\Rr^{+}$ is
$\Rr-\{0\}$, $R$ is the real interval [0,1], $Q$ is the
set of rational numbers in $R$, $B$ is the set of numbers in $R$ 
with finite radix-2 representations,  $X =$ \{0,1\}, $X^*$ is the set
of finite binary sequences including the null sequence
\Nil, and $X^{\oo}$ is the set of infinite binary sequences.
$|x|$ is the length (number of digits) of a sequence $x \in X^*$.
If $|x|=n$ then $x$ is said to be an $n$-sequence.
The notation $x \px y$ (or $y \xt x$) expresses that $x$ is a
prefix of $y$ ($y$ is an extension of $x$), where $x \in\ X^*$ and 
$y \in X^*\cup X^{\oo}$. Similarly $x \ppx y$ (or $y \pxt x$) expresses
that $x$ is a proper prefix of $y$, i.e., $x \px y$ and $x \neq y$.
A prefix of length $n$ of a finite or infinite sequence $x$ is 
denoted by $x(n)$ ($x(n)$ is undefined if $n>|x|$).
The concatenation of two sequences
$x$ and $y$ is written as $xy$. Similarly \{$xy|x\in S, y\in T$\}
is written as $ST$. Set concatenations involving singletons
(e.g., \{$x\}X^*$) are shortened by omitting braces of the 
singletons (e.g., $xX^*$). Also $n$-fold self-concatenation of
a sequence $x$ or set of sequences $S$ is written as
$x^{n}$ or $S^{n}$ respectively.
Note that $\emp S = S\emp = \emp$. A set $S \pss X^*$
is \UL{prefix-free} (pf) iff $S\cap (SXX^*) = \emp$, i.e., it
contains no proper extension of any of its members.
Such a set is also called an \UL{instantaneous code}
(Abramson, 1963).

A \UL{predictor} is a total function $p: X^*\ra R$ such that
$p(x) \geq p(x0)+p(x1)$ for all $x\in X^*$. Thus $p$ corresponds
either to a subadditive measure $\mu$ on $X^{\oo}$ such that
$\mu xX^{\oo}$ = $p(x)$ or to an additive measure $\mu$ on 
$X^*\cup X^{\oo}$ such that $\mu x(X^*\cup X^{\oo})$ = $p(x)$, for all 
$x \in X^*$. Intuitively $p(x)$ may be regarded as the prior probability
of $x$ or as a lower bound on its prior probability (see below).
The difference $p^{+}(x)$ = $p(x) - p(x0) - p(x1)$ is called 
the \UL{surplus} probability of $x$.
A predictor satisfying $p(\nil) = 1$ and $p(x) = p(x0)+p(x1)$
for all $x \in X^*$ is called a \UL{sequential probability
distribution} (Martin-L\"{o}f, 1966), or a distribution,
for short.\footnote
  {Solomonoff (1964) used the term ``normalized
  probability evaluation methods" for computable 
  distributions. Schnorr (1971) defined randomness
  in terms of martingales, where a martingale 
  $f:X^*\ra \Rr^{+}$ satisfies $f(x)\,= (f(x0)+f(x1))/2$
  for all $x\in\,X^*$. Thus if $f(x)\leq 2^{|x|}$ for all $x \in X^*$,
  then $2^{-|x|}f(x)$ defines a sequential probability distribution.
  See Sec.\,4.}

 A total function $f:X^*\ra R$ is \UL{incrementable} iff there
is a recursive function $g:\,X^*\x\,N\,\ra\,Q$ which is nondecreasing 
in its second argument such that\\
 \ind $f(x) = \limn g(x,n)$ for all $x \in X^*$;\\
i.e., $f(x)$ is approachable from below, with each
increase in $n$ supplying a nonnegative rational increment
in the approximation $g(x,n)$ to $f(x)$.\footnote
  {It is assumed that procedures which accept rational 
  numbers as inputs or generate them as outputs utilize 
  some effective encoding of the rational numbers, e.g.,
  integer pairs $\lc m,n\rc$ such that $m/n = q$. Instead of the
  rational numbers a more restricted set such as $B$ (the numbers
  with finite radix-2 representations), or a less restricted
  set such as the computable numbers in $R$ could have been used.}
When a recursive function $g$ and a function $f$ are related
as above, $g$ is said to \UL{underlie} $f$.

One of the primary concerns in this paper will be
the class of incrementable predictors.\footnote
  {These correspond exactly to Levin's semicomputable measures
  (Zvonkin \& Levin, 1970), apart from the inessential condition $p(\nil)$
  = 1, i.e., $\mu X^*\cup X^{\oo}$ = 1, on any semicomputable measure \Mu.}
The importance of this class of predictors lies in its relationship
to the class of Martin-L\"{o}f random sequences on the one
hand (Sec.\,3) and to the class of processes on the other (Sec.\,5).

\noind
The following two simple facts about predictors are noteworthy.

\noind
{\bf Theorem 1.} (a) Every incrementable distribution is recursive.\\
(b) Every recursive predictor can be increased to a recursive 
distribution.

\noind
\UL{Proof.} (a) If $p$ is any incrementable distribution
then $p(\nil)$ is trivially computable.
Assume for induction on $n$ that $p(x)$ is computable 
for every $n$-sequence $x$. Now from $p(x0) = p(x)-p(x1)$ it is seen
that $p(x0)$ is approachable from above, since $p(x)$ is computable
by assumption and $p(x1)$ is approachable from below.
But $p(x0)$ is also approachable from below, so that 
$p(x0)$ is computable; similarly for $p(x1)$.\\*[.03in]
(b) For any recursive predictor $p$, a distribution $p'$
such that $p'(x)\geq p(x)$ for all $x \in X^*$ can be defined as follows:
Let $p'(\nil) = 1, p'(x0) = p'(x) - p(x1)$, and $p'(x1) = p(x1)$
for all $x \in X^*$. Then it is easily verified by induction
on sequence length that p$'$ meets the requirements of the
theorem. $\Box$

 In what sense and under what conditions does
a predictor allow sequence prediction? If the predictor 
is a recursive distribution the answer is straightforward.
Consider any \UL{nonterminating} process which
generates a succession of binary digits; then $p(xy)/p(x)$ 
can be regarded as the conditional probability that
$x$ will be followed by $y$, given that $x$ has occurred
(replace any ratio 0/0 by 0). Thus a ``prediction" of
a sequence continuation is analogous to a weather
forecast, say, which attributes a probability to some
future weather condition (e.g., ``60\% chance of rain
tomorrow").\footnote
  {The ``rationality" of such predictions depends on the 
  frequency with which events assigned particular conditional
  probabilities occur; see discussion preceding Th.\,11.}

 Arbitrary predictors, however, admit two
intuitive interpretations, corresponding to the two
measure-theoretic interpretations mentioned above.
In the first interpretation a predictor supplies {\it lower
bounds} on prior probabilities of initial output sequences
generated by a {\it nonterminating} process. Corresponding
upper and lower bounds on conditional probabilities are
supplied in Th.\,2. These are approachable from above and below 
respectively, whenever the given predictor is incrementable (Th.\,3).

 In the second interpretation a predictor supplies prior
probabilities on nonempty output sequences of a process {\it which
may or may not terminate}. The surplus probability of
a sequence $x \in XX^*$ is then the probability that the process will
generate $x$ and halt. As in the case of distributions, 
$p(xy)/p(x)$ is the probability that $y$ will follow $x$, given
that $x$ has been generated, but with no guarantee that a
continuation of length $|y|$ will be generated at all.
These conditional probabilities need not be approachable
from below, even if the given predictor is incrementable (Th.\,4).

 The following theorem gives the sharpest possible bounds
on conditional probabilities implicit in the values
of a predictor, when these are interpreted as lower bounds
on initial-segment probabilities in a nonterminating process
(first interpretation). For any $x \in XX^*$, $x^-$ denotes the
sequence obtained by changing the last digit of $x$ to 
its complement. Thus $v(i)^-$ is $v(i)$ with the $i$th 
digit complemented. A sum over no terms (in particular,
a sum from a higher to a lower summation index) is 
taken to be 0. As before, occurrences of 0/0 are to
be replaced by 0.

\noind
{\bf Theorem 2.} If $p$ is any predictor and $p'$ is any 
distribution such that $\al y\in X^*:\,p'(y)\geq p(y)$, then
$\al v\in X^*:\,\al w\in XX^*$:

\[ \frac{p(vw)}{1 - \sum_{i=1}^{|v|}{p(v(i)^-)}} \leq
  \frac{p'(vw)}{p'(v)} \leq 
  \frac{1 - \sum_{i=1}^{|vw|}{p((vw)(i)^-)}}{1 - \sum_{i=1}^{|v|}{p(v(i)^-)}}. \]

\noind
Furthermore, these are the sharpest possible bounds
derivable from $p$ in the sense that $\al v\in X^*:\,\al w\in XX^*$:
\Ex\ distributions $p'$, $p''$: $\al x\in X^*:\;p'(x), p''(x) \geq p(x)$ and 

\[ \frac{p'(vw)}{p'(v)} = \frac{p(vw)}{1 - 
                                \sum_{i=1}^{|v|}{p(v(i)^-)}} \ \ {\rm and} \ \
  \frac{p''(vw)}{p''(v)} = \frac{1 - \sum_{i=1}^{|vw|}{p((vw)(i)^-)}}
                                {1 - \sum_{i=1}^{|v|}{p(v(i)^-)}}. \]
\noind
\UL{Proof.} The lower bound on $p'(vw)/p'(v)$ is obtained
from the lower bound $p(vw)$ on $p'(vw)$ and upper bound
\( p'(\nil) = \sum_{i=1}^{|v|}{p(v(i)^-)} \)

\noind
on $p'(v)$, which is easily inferred from the distribution property
of $p'$ and the fact that each $p(v(i)^-$) is a lower bound 
on $p'(v(i)^-)$, for $i = 1,...,|v|$. The upper bound on 
$p'(vw)/p'(v)$ is obtained by noticing that the difference
between $p'(v)$ and $p'(vw)$ is at least\\*[.03in]
\ind \( \sum_{i = |v|+1}^{|vw|}{p((vw)(i)^-)}, \)\\*[.03in]
so that $p'(vw)/p'(v)$ is maximized by choosing $p'(vw)$ as
large as possible while keeping $p'(v)-p'(vw)$ to its minimum.

The second part of the theorem is proved by constructing $p'$ such that\\*[.03in]
\ind\( p'(x) = 1 - \sum_{i=1}^{|x|}{p(x(i)^-)}, \) \ and\\
\ind\( p'(x^-) = p(x^-) \) \  for~all $x \ppx\,vw$;\\*[.03in]
\ind\( p'(vw) = p(vw) \); \ and\\
\ind\( p'(vw^-) = 1 - \sum_{i=1}^{|vw|-1}{p((vw)(i)^-)} - p(vw). \)
     
\noind
Then $p'$ is easily seen to be a distribution bounded below
by $p$ for all $x$ and $x^-$ such that $x\px\,vw$; its 
extension to other $x \in X^*$ is straightforward.
Evidently $p'(vw)/p'(v)$ equals the lower bound of the
theorem. Similarly $p''$ is constructed such that\\*[.03in]
\ind\( p''(x) = 1 - \sum_{i=1}^{|x|}{p(x(i)^-)} \), \ and\\*[.03in]
\ind\( p''(x^-) = p(x^-) \ {\rm for~all} \ x \ \px vw. \)

\noind
Again $p''$ is easily seen to be a distribution bounded
below by $p$ for all $x$ and $x^{-}$ such that $x\px vw$; its
definition is easily completed. Evidently $p''(vw)/p''(v)$
equals the upper bound of the theorem.~$\Box$

\noind
{\bf Theorem 3.} If the predictor $p$ of Th.~2 is
incrementable, then the upper and lower bounds on
conditional probability of that theorem are approachable
from above and below respectively.

\noind
\UL{Proof.} Since $p$ is approachable from below, the
given lower bound on $p'(vw)/p'(v)$ is approachable from
below. Inspection of the upper bound indicates that whenever
$p(x^{-})$ is incremented for some $x\px vw$, the numerator
will decrease while the denominator will either decrease by the
same amount or remain unchanged. In neither case does the
upper bound increase, since the numerator is at most as
large as the denominator.~$\Box$

 Thus the conditional predictor determined by
an incrementable predictor under the first interpretation is
itself incrementable. This is not the case under the second
interpretation, i.e., the probability $p'(x,y)$ that $y$ follows
$x$ is not approachable from below, as the following theorem 
shows. The theorem also deals with the non-incrementability,
under the second interpretation of predictors,
of the probability $p''(x,y)$ that $y$ follows $x$, given
that the process does {\it not} terminate prematurely (i.e.,
after generating $xy'\ppx\,xy)$. This probability would be of interest
if it were known that a continuation of length $\geq~|y|$ had
been generated, but its digits were still unknown.
Solomonoff's (1964, 1976) and Cover's (1974) conditional
probabilities are of this type.

\noind
{\bf Theorem 4.} If $p$ is a predictor then p$'$ defined as\\   
 \ind $p'(x,y) = p(xy)/p(x)$ for all $x, y\in\,X^*$\\
and $p''$ defined as\\
 \ind $p''(x,\nil) = p(x)/p(x)$,\\
 \ind $p''(x,w) = p(xw)/(p(xw)+p(xw^{-}))$, and\\
 \ind $p''(x,yw) = p''(x,y)\,p''(xy,w)$ for all $x, y \in X^*$ and $w \in X$,\\
are not in general approachable from below.

\noind
\UL{Proof.} Let $\phi_i:\,N\ra\,N, i = 0, 1, 2, ...$ be a
recursive enumeration of the partial recursive functions.  Let\\*[.01in]
\ind $p(0^m) = 2^{{\rm-min}\{n| n\geq m, \ \phi_n \ {\rm defined}\}}$ for all
                                                             $m\in N$, and\\
\ind $p(x) = 0$ for all $x\in X^*1X^*$.\\*[.03in]
Clearly $p$ is a predictor. Since $\{n| \phi_n(n) \ {\rm defined}\}$
is recursively enumerable (re), $p$ is incrementable. It is
easily seen that $p(0^{m+1})/p(0^m) > .5$ iff $\phi_{m}(m)$
is undefined. But $p(0^{m+1})/p(0^m) = p'(0^{m},0)$.
Hence if $p'$ were approachable from below one could eventually
verify that $p(1^{m+1})/p(1^{m}) > .5$ whenever this is the case.
But then one could recursively enumerate $\{m| \phi_m(m) \ {\rm
undefined}\}$, which is well-known to be impossible (e.g.,
see Rogers, 1967). Hence $p'$ is not approachable from below.

\noind
 Now let $p$ be redefined as follows:\\
\ind $p(0^m1) = 2^{-m-1}$ if $\phi_m(m)$ is defined,\\
\hspace*{.65in} = 0 otherwise, for all $m\in N$;\\
\ind $p(0^m) = \sum_{i=0}^{\oo} p(0^{m+i}1)$ for all $m\in N$; and\\
\ind $p(x) = 0$ for all $x\in X^*-\{0\}^*\{\nil,1\}$.

\noind Again $p$ is easily shown to be an incrementable
predictor, with the property that $\al m\in N$:\\*[.03in]
\ind $p(0^{m+1})/(p(0^{m+1}) + p(0^{m}1) > .5 $
                                    iff $\phi_{m}(m)$ is undefined.\\
But this ratio is $p''(0^{m},0)$ so that if $p''$ were approachable
from below, $\{m|\,\phi_{m}(m) \ {\rm undefined}\}$ would be re.
Hence $p''$ is not approachable from below.~$\Box$

 This result indicates that some incrementable predictors
are far from ``effective" as methods of prediction, particularly
under the second interpretation of predictors. Therefore
it is important to study narrower classes of predictors,
such as the recursive predictors (Sec.~4).

\subsection*{3. Quasipredictability and Martin-L\"{o}f Randomness}
 The \UL{redundancy} of a sequence $x\in\,X^*$ relative to
a predictor $p$ is defined as $rp(x) = |x| + \log~p(x)$,
with logarithms taken to base 2 and $\log 0 = -\oo$.
This can be thought of as the maximum possible information of $x$,
viz.~$|x|$, less (the upper bound on) its actual information, viz.\,$-\log p(x)$.
For a sequence $z\in\,X^{\oo}$ the redundancy relative to $p$
is defined as $rp(z) = \limsupn\,rp(z(n))$. Note that every sequence
has zero redundancy relative to the uniform distribution
$p(x) = 2^{-|x|}$.

 An important related notion is that of optimality. 
A predictor is optimal for a given infinite sequence
if it ``reveals the regularities (redundancies)" of that
sequence essentially as well as any other predictor.
Formally, an incrementable predictor $p$ is \UL{optimal for
$z$} $z\in\,X^{\oo}$, iff $rp'(z(n)) - rp(z(n))$ is 
bounded above for every incrementable predictor $p'$.

 A sequence $z \in X^{\oo}$ is \UL{quasipredictable}
iff its redundancy is \Oo\ relative to some incrementable
predictor. The prefix ``quasi" indicates that prediction
with an incrementable predictor is not fully effective.

 Quasipredictability will now be related to Martin-L\"{o}f
randomness.

 A \UL{Martin-L\"{o}f (M-L) sequential test} is a set
$V \pss\,X^*\x\,N$ with the following 4 properties,
where $V_{m}$ denotes $\{x|\,\lc x,m\rc\in\,V\}$:\\
\ind (a) Effectiveness: $V$ is re.\\
\ind (b) Nestedness: $V_{m+1}\,\sseq\,V_{m}$ for all $m\in\,N$.\\
\ind (c) Numerosity: the number of $n$-sequences in $V_{m}$ $\leq 2^{n-m}$ 
                                                 for all $m, n \in N$.\\
\ind (d) Monotonicity: $x\in\,V_{m}$ \RA\, $xy\in\,V_{ m}$ 
                                          for all $x, y \in X^*$.\\*[.03in]
For motivation of these properties see Martin-L\"{o}f (1966).\footnote
  {Alternative definitions can be found in Zvonkin \& Levin (1970)
  and Schnorr (1971, 1973).}
Intuitively, $\lc x,m\rc\,\in\,V$ means that the test $V$ rejects the
randomness hypothesis at significance level $2^{-m}$ for all
infinite sequences beginning with $x$.

 The \UL{critical level} $m_{V}(x)$ of a sequence $x\in\,X^*$ relative to
a M-L sequential test $V$ is max$\{m|\,\lc x,m\rc\,\in\,V\}$, where max\,\Emp = 0
(this latter condition in effect extends $V_{0}$ to $X^*$).
A sequence $z$ \In  $X^{\oo}$ is \UL{M-L random} iff there is no 
M-L sequential test $V$ such that \Limn $m_{V}(z(n)) = \oo$.

 More generally, a \UL{M-L sequential test for $p$} where
$p$ is any recursive distribution, is defined as above except
that the numerosity condition becomes \\*[.03in]
\ind \( \sum_{y\in X^{n}\cap V_{m}} p(y) \leq 2^{-m} \)
                                             for all $m, n \in N$.%
\footnote
  {Martin-L\"{o}f originally used strict inequality to facilitate
  the construction of a universal test (because equality of
  computable reals is not effectively confirmable). However,
  extension of $V_{0}$ to $X^*$ is then no longer possible,
  and in any case the construction of a universal test
  is still possible with non-strict inequality
  (see Zvonkin \& Levin, 1970).}\\*[.03in]
Accordingly a sequence $z\,\in\,X^{\oo}$ is \UL{M-L random relative
to $p$} where $p$ is any recursive distribution, iff there is
no M-L test $V$ for $p$ such that $\limn\,m_{V}(z(n)) = \oo$.
Note that ``M-L random" is the same as ``M-L random relative to
the uniform distribution".\footnote
  {It should be mentioned that the notion of randomness relative
  to recursive distributions does not allow for randomness relative
  to Bernoulli distributions of the form $p(\nil) = 1, p(x0) = 
  p(x)(1-r), p(x1) = p(x)r$ for any (not necessarily computable)
  $r\in R$. For a treatment of Bernoulli sequences see Martin-L\"{o}f
  (1966) and Schnorr (1971). For randomness tests relative to
  arbitrary distributions see Levin (1976).}

 Intuitively, one would expect that if a distribution is optimal
for a sequence $z\,\in\,X^{\oo}$, i.e., if it reveals all the
regularities (redundancies) of $z$, then $z$ should appear to behave 
randomly relative to the probability assignments of the distribution.
This is indeed the case.

\noind
{\bf Theorem 5} (Levin). For any recursive distribution $p$ and
any $z\,\in\,X^{\oo}$, $z$ is M-L random relative to $p$ iff $p$ is
optimal for $z$.

\noind
\UL{Proof.} \RA: Suppose that $rp'(z(n)) - rp(z(n))$, i.e., 
$\log p'(z(n)) - \log p(z(n))$, is unbounded for some incrementable
predictor $p'$. Let\\*[.03in]
 \ind $V = \{\lc x,m\rc |\,x\in\,X^* \ \& \ m\in\,N \ \& \
                              \ex y\px\,x: p'(y)>2^{m}p(y)\}$.\\*[.03in]
$V$ will be shown to be a M-L test for $p$ such that $z$ is not
M-L random relative to $p$.

 Nestedness and monotonicity are obviously satisfied by $V$.
$V$ is clearly re since $p'$ is incrementable and $p$ is recursive.
The numerosity condition can be verified by considering a
partitioning of the $n$-sequences in $V_{m}$ into groups such
that group $i$ consists of all $n$-sequences extending some $y_{i}$
with $p'(y_{i}) > 2^{m}p(y_{i})$. Then

\begin{eqnarray*}
   1 \geq \sum_i p'(y_i)  >  2^{m}\sum_i p(y_i)
                              = 2^{m}\sum_{\stackrel{|x|=n}{x\xt y_i}} p(x)
                         =  2^{m}\sum_{\stackrel{|x|=n}{x \in V_m}} p(x),\\
   \mathrm{so \ that} \ \ 
  2^{m}\sum_{\stackrel{|x|=n}{x \in V_m}} p(x) < 2^m.
\end{eqnarray*}

Now from the definition of $V$, $\lc z(n),m\rc\,\in\,V$ if $\log\,p'(z(n))$
$> m + \log p(z(n))$. But $\limsupn [\log p'(z(n)) - \log p(z(n))]$
$= \oo$, hence $\limn m_{V}(z(n)) = \oo$ and $z$ is not M-L random
relative to $p$.\\
\LA: Assume without loss of generality that $p(z(n))$ does not 
vanish for any $n$ (otherwise $\log p'(z(n)) - log~p(z(n))$ will
certainly be unbounded for any nonvanishing $p'$). Let $V$ be a
M-L test for $p$ such that $\limn m_{V}(z(n)) = \oo$. Let $p'$(x) be
defined for all $x \in X^*$ as\\*[.03in]
 \ind\ind $p'(x) = \limn q(x,n)$, where\\
 \ind\ind \( q(x,n) = \sum_{\stackrel{y\xt x}{|y|=n}} m_V(y)p(y) \) \ 
                                                   for $n > |x|$.\\*[.03in]
$p'$ will be shown to be an incrementable predictor such that
$\log p'(z(n)) - \log~p(z(n))$ is unbounded. First observe that
q(x,n) is nondecreasing in $n$, as $m_{V}(y0), m_{V}(y1)$
$\geq m_{V}(y)$, so that

\begin{eqnarray*}
 \sum_{\stackrel{y\xt x}{|y|=n+1}} m_V(y)p(y) & = & 
                        \sum_{\stackrel{y\xt x}{|y|=n}}[m_V(y_0)p(y0) 
                                                       + m_V(y1)(p(y1)] \\
                & \geq & \sum_{\stackrel{y\xt x}{|y|=n}} m_V(y)p(y).
\end{eqnarray*}
Furthermore $q(x,n)\leq q(\nil,n)\leq 1$ for all $x, n$ since

\begin{eqnarray*}
 q(\nil,n) & = & \sum_{|y|=n}~m_V(y)p(y) = \sum_{m\in N} m 
                                            \sum_{\stackrel{|y|=n}{m_V(y)=m}}p(y) \\
    &  = & \sum_{m\in N}m\left[ \sum_{y\in X^n\cap V_m}p(y) 
                                          - \sum_{y\in X^n\cap V_{m+1}}p(y)\right] \\
    & = & \sum_{m=1}^{\oo} \ \sum_{y\in X^n\cap V_m} p(y) 
                                          \leq \sum_{m=1}^{\oo} 2^{-m} = 1.
\end{eqnarray*}
\noind
Thus $\limn\,q(x,n)$ exists and is $\leq 1$ for all $x$. From
$q(x,n) = q(x0,n) + q(x1,n)$ for all $n > |x|$, it follows
that $p'(x) = p'(x0) + p'(x1) \geq 0$ so that $p'$ is a predictor
(in fact, an additive predictor). From the fact that $V$ is re
and that for all $x \in X^*$ and $r \in R$\\*[.03in]
\ind\ind $p'(x) > r$ \ \RA \ $\ex n: q(x,n) > r$\\*[.03in]
 it is easy to see that $p$ is incrementable. Now since
$q(x,n)$ is nondecreasing in $n$,\\*[.03in]
\ind\ind  $p'(z(n))\geq q(z(n),n) = m_{V}(z(n))p(z(n))$.\\*[.03in]
Hence if $m_{V}(z(n))$ is unbounded,
so is $p'(z(n))/p(z(n))$ and hence also $\log p'(z(n)) - \log p(z(n))$. $\Box$

\noind
\UL{Remark.} It would have been possible to use
 \[ q(x,n) = \sum_{\stackrel{y\xt x}{|y|=n}} f(m_V(y))p(y) \]
in the proof, where $f$ is any unbounded nondecreasing recursive
function from $N$ to \RR\ such that
\[ \sum_{m\in N}{f(m) 2^{-m-1}} \leq 1; \]
the reason is that $q(\nil,n)$ then becomes
\begin{eqnarray*}
 \lefteqn{
  f(0) + \sum_{m=1}^{\oo}\left[ (f(m) - f(m-1)) 
                                   \sum_{y\in X_n\cap V_m} p(y)\right]}\\
   & & \leq f(0) + \sum_{m=1}^{\oo}{(f(m) - f(m-1)) 2^{-m}} 
                                               = \sum_{m=0}^{\oo} 2^{-m-1}.
\end{eqnarray*}

 A particularly interesting corollary of Th.~5 results from
specializing $p$ to the uniform distribution. This provides the
link between quasipredictability and M-L randomness.

\noind
{\bf Corollary 1.} A sequence $z$ \In  X$^{\oo}$ is M-L random iff
it is not quasipredictable.

In other words, the M-L random sequences are those for 
which the uniform predictor is optimal.

 The next corollary supplies a predictor which is \UL{universal}
in the weak sense that it assigns infinite redundancy to all
nonrandom sequences. A predictor which is optimal for all sequences
(and hence certainly universal) is given in Th.\,8.

\noind
{\bf Corollary 2} (universal predictor). There is
an incrementable predictor $p$ such that for any $z\in X^{\oo}$
and any incrementable predictor $p'$\\*[.03in]
\ind\ind $rp'(z) = \oo$ \ \RA\ \ $rp(z) = \oo.$\\*[.03in]
A suitable $p$ is given by 
\[ p(x) = \limn 2^{-n} \sum_{\stackrel{y\xt x}{|y|=n}} m_U(y), \]
where $U$ is a universal M-L sequential test (Martin-L\"{o}f, 1966).

\noind
{\bf Corollary 3.} If $p$ is a universal predictor then
$p(x) > 0$ for all $x\in X^*$.

\noind
\UL{Proof.} For any $x \in X^*$, consider the sequence $z = xy$,
where $y\in X^{\oo}$ is some fixed recursive sequence.
Clearly there is an incrementable predictor $p'$ such that
$p'(z(n)) = 1$ for all $n\in  N$. Hence $z$ is nonrandom and
hence $rp(z) = \oo$. But if $p(x)$ were 0, $rp(z(n))$
would be $-\oo$ for all $n\geq |x|$. $\Box$

 Universal predictors are
not recursive. This is the analogue of the fact that
there is no recursive universal M-L sequential test.

\noind
{\bf Theorem 6.} There is no recursive universal predictor.

\noind
\UL{Proof.} By Cor.~3 of Th.~5 it is only necessary to prove
that any nonvanishing recursive predictor $p$ is not universal.
Now if $p$ is recursive then arbitrarily tight upper and lower
bounds on $p(x)$ can be effectively computed for any $x \in X^*$.
This fact allows the construction of a recursive sequence
$y\in X^{\oo}$ whose redundancy is bounded relative to $p$,
showing that $p$ is not universal. Specifically, the $(n+1)$st
digit of $y$ is chosen as follows. Increasingly tight upper
and lower bounds on $p(y(n)), p(y(n)0), \ {\rm and} \ p(y(n)1)$ are
computed. $y(n+1)$ is assigned the value $y(n)0$ if the inequality\\*[.03in]
\ind\ind \( p(y(n)0) < p(y(n))2^{2^{-n}-1}, \)\\*[.03in]
is first confirmed, or the value $y(n)1$ if\\*[.03in]
\ind\ind \( p(y(n)1) < p(y(n))2^{2^{-n}-1} \)\\*[.03in]
is first confirmed. One of these inequalities will be confirmed
eventually since the multiplier of $p(y(n))$ on the right-hand side
exceeds 1/2 for all $n\in N$, and $p(y(n)0), p(y(n)1)$ cannot
both exceed $p(y(n))/2$. Thus\\*[.03in]
\ind\ind \( p(y(n+1)) < p(y(n))2^{2^{-n}-1} \)\\*[.03in]
for all $n$ and hence
\[ p(y(n)) < p(\nil) \prod_{i=0}^{n-1}{2^{2^{-i} - 1}} 
                                        = p(\nil) 2^{2-2^{-n+1} -n}. \]
\noind
Consequently the redundancy of $y$ satisfies\\*[.03in]
\ind\ind $n + \log P(y(n)) < \log p(\nil) + 2 - 2^{-n+1}$\\*[.03in]
and so is bounded above. Thus $p$ is not universal. $\Box$

\noind
{\bf Corollary.} There is no incrementable universal distribution.

\noind
\UL{Proof.} Immediate from Th.~1(a) and Th.~6. $\Box$

\noind
Note that this implies that for any additive universal
predictor $p(\nil)$ is not computable.

 The predictor of Cor.~2, Th.~5 was seen to be universal
in that it assigns infinite redundancy to all infinite
nonrandom sequences. However, it is nonoptimal, i.e., 
it does not in general maximize
initial-segment redundancy. The fact that
any universal M-L sequential test maximizes the initial-segment
critical level of any infinite sequence, apart from a constant
(Martin-L\"{o}f, 1966), suggests that it should be possible to
derive an optimal predictor from such a M-L test.
The following theorem does not quite
succeed in confirming this intuition. The predictor
exhibited falls short of maximizing redundancies by a term
logarithmic in the redundancy. Its special interest lies 
in the fact that it is additive. In Th.~8 the existence of
a truly optimal (but nonadditive) universal predictor will be
established without reliance on the properties of M-L tests.

\noind
{\bf Theorem 7.} There is an additive incrementable 
predictor $p$ such that for any incrementable predictor $p'$
and any real $c>1$\\*[.03in]
\ind\ind $rp'(x) - rp(x) < c \log rp'(x)$\\*[.03in]
for all $x \in X^*$ such that $rp'(x)$ is sufficiently large
(i.e., larger than some constant dependent on $p'$ and $c$).
A suitable $p$ is given by
\[ p(x) = \limn 2^{-n}\sum_{\stackrel{y\xt x}{|y|=n}} f(m_U(y)), \]
\noind
where $\log f(m) = m - \log(m+2) -2\log\log(m+5)$ for all $m\in N$,\\*[.03in]
and $U$ is any universal M-L sequential test.

\noind
\UL{Proof.} It can be verified  that $f$ is nondecreasing  and satisfies\\*[.03in]
\ind\ind \( \sum_{m\in N}{f(m) 2^{-m-1}} \leq 1. \)\\*[.03in]
By the proof of Th.~5 and the remark following it, $p$ is
an additive incrementable predictor. Given any incrementable
predictor $p'$, let\\*[.03in]
\ind\ind $V = \{\lc x,m\rc | x\in X^* \ \& \ m\in N \ \& \ 
                                      \ex y\px x: p'(y) > 2^{m-|y|}. \}$\\*[.03in]
For this M-L sequential test it is known that if $\log p'(x) > m-|x|$
then $m_{V}(x)\geq m$. Also, since $U$ is universal
there is an integer $a$ such that $m_{U}(x)\geq m_{V}(x)-a$
for all $x \in X^*$. Clearly $p(x)\geq 2^{-|x|}f(m_{U}(x))$
or,\\*[.03in] 
\ind $\log p(x)\geq -|x|+m_{U}(x)-\log(m_{U}(x)+2)-2\log\log(m_{U}(x)+5)$.\\*[.03in]
Thus $\log p'(x) > m-|x|$ implies $m_{U}(x)\geq m-a$ which
in turn implies\\*[.03in]
\ind $\log p(x)\geq -|x|+m-a-\log(m-a+2)-2\log\log(m-a+5)$\\*[.03in]
provided that $m-a\geq 0$ (in view of the fact that $f(i)$ is 
nondecreasing in $i$ for $i\in  N$); or, $|x|+\log p'(x) > m\geq a$
implies\\*[.03in]
\ind  $|x|+\log p(x)\geq m-[a+\log(m-a+2)+2\log\log(m-a+5)]$;\\*[.03in]
or, choosing $m$ such that $m+1\geq |x|+\log p'(x) \ > \  m$ and 
assuming that $|x|+\log p'(x) > a$,\\*[.03in] \ind $\log p'(x)-log~p(x)$\\*[.03in] 
\ind $< 1 + a + \log(|x|+\log p'(x)-a+2) + 2\log\log(|x|+\log p'(x)-a+5)$\\*[.03in]
\ind $ < c\log rp'(x)$ for all sufficiently large $rp'(x)$\\*[.03in]
and the theorem follows. $\Box$

 Essentially the following lemma says that the class of
incrementable predictors is re,
and that the function underlying a predictor can be chosen
to be subadditive (like the predictor itself). This fact
facilitates the construction of an optimal predictor (Th.~8)
and the reduction of predictors to Solomonoff's form (Th.~12).

\noind
{\bf Lemma 1.} There is a re class of recursive functions
$\{h_{i}: X^*\x N \ra Q, \ i\in N\}$ each of whose members
$h_{i}$ underlies a predictor $p_{i}$ and satisfies\\*[.03in]
\ind\ind $h_{i}(x,n)\geq h_{i}(x0,n) + h_{i}(x1,n)$\\*[.03in]
for all $x \in X^*$, $n\in N$. Furthermore $\{p_{i}| i\in N\}$
is the class of all incrementable predictors.

\noind
\UL{Proof.} It will first be shown that the class of
incrementable {\it functions} is re. From any partial
recursive function $\phi: X^*\x N \ra Q$ a recursive function
$g: X^*\x N \ra Q$ can be obtained uniformly effectively, such
that $\lm n\,g(x,n)$ is nondecreasing and \\*[.03in]
\ind\ind $\limn\,g(x,n) = \sup{\phi(x,n)| n\in N}$
for all $x \in X^*$, where $\sup \emp = 0$.\\*[.03in]
 Given a procedure
for \PHI, $g(x,n)$ can be computed by simulating the
computation of $\phi(x,0),...,\phi(x,n)$ for $n$ steps each
and returning the largest of the outputs obtained (0 if
none are obtained). If \PHI\ is already a recursive 
nondecreasing function, then $\limn g(x,n)$ will clearly be
the same as $\limn\phi(x,n)$. Since the class of partial
recursive functions is re, it follows that the class of
incrementable functions is re.

The members of this class can now be further
modified to yield only (and all) incrementable predictors.
For any incrementable $g$, let\\*[.03in]
 (1) $h(x,0) = 0$,\\
 (2) $h(\nil,n) = g(A,n)$\\
 (3) $h(x0,n) = \min\{g(x0,n), h(x,n)-h(x1,n-1)\}$ if $n>0$, and\\
 (4) $h(x1,n) = \min\{g(x1,n), h(x,n)-h(x0,n)\}$,
 for all $x \in X^*$, $n\in  N$.\\
Assume for induction on $|x|$ that $h(x,n) \geq h(x,n-1)$ for
all $n>0$ and all $x$ of length $\leq k$. Then the inductive
step requires proving $h(x0,n) \geq h(x0,n-1)$ and $h(x1,n)
\geq h(x1,n-1)$, i.e., \\*[.03in]
 (5) $g(x0,n) \geq h(x0,n-1)$,\\
 (6) $h(x,n)-h(x1,n-1) \geq h(x0,n-1)$,\\
 (7) $g(x1,n)\geq h(x1,n-1)$, and\\
 (8) $h(x,n)-h(x0,n) \geq h(x1,n-1)$.\\
Since $g$ is nondecreasing and since $g(x,n) \geq h(x,n)$
for all $x$, $n$ (by inspection of (1)-(4)), therefore
(5) and (7) hold. From (4)\\*[.03in]
 (9) $h(x,n-1) \geq h(x0,n-1)+h(x1,n-1)$\\
for all $x$ and $n>0$, and together with the induction
assumption this implies (6). (8) is immediate from (3).
The basis of the induction is $h(\nil,n) = g(\nil,n) \geq
g(\nil,n-1) = h(\nil,n-1)$. Thus $h$ is nondecreasing and hence
(9) holds in the limit as $n\ra \oo$, i.e., $h$ underlies a 
predictor.

 Whenever $g$ underlies a predictor $p$, $h$ underlies
the same predictor. This is proved by assuming for induction
that for all $x$ of length $\leq k$ and all $r\in R$, if 
$\ex n\in N: g(x,n) > r$ then $\ex n\in N: h(x,n) > r$.
Suppose that $\ex r\in R: \ex n\in N: g(x0,n) > r$. Then
since $g$ underlies predictor $p$, $\ex n'\in N: g(x,n') > r+p(x1)$, 
and hence $\ex n''\in N: h(x,n'') > r+p(x1)$, by the induction
assumption. Hence by (3) either $h(x0,n'') = g(x0,n'') > r$ 
(assuming w.l.g. that $n''\geq n$) or $h(x0,n'') = h(x,n'')-h(x1,n'-1)
> r+p(x1)-g(x1,n''-1)\geq r$. The argument for $h(x1,n)$ is similar,
and equation (2) starts the induction. Since the transformation
from $g$ to $h$ is uniformly effective, and $h$ is subadditive (see
equation (9)), the proof of the lemma is complete. $\Box$

 The construction of the following optimal predictor is
modelled on Martin-L\"{o}f's (1966) costruction of a universal
test. An alternative construction follows as an easy corollary
of the reduction of incrementable predictors to Solomonoff
predictors (see corollary of Th.~12); this was the approach
used by Levin (in Zvonkin \& Levin, 1970). However, the required
reduction is itself nontrivial, so that an approach not 
dependent upon it is of some interest. 

\noind
{\bf Theorem 8} (Levin, optimal universal predictor). There is an
incrementable predictor $p$ such that for any incrementable $p'$ 
there is a constant $c$ satisfying \\*[.03in]
\ind\ind $rp'(x) - rp(x) \leq c$ for all $x \in X^*$.

\noind
\UL{Proof.} With $p_{i}$, $i\in N$, defined as in Lemma~1,
the optimal universal predictor is given by\\[.03in]
\ind\ind \(p(x) = \sum_{i=0}^{\oo} 2^{-i-1} p_i(x)\) for all $x\in X^*$.\\[.03in]
Then $p(x) = \limn h(x,n)$ for all $x\in X^*$, where\\[.03in]
\ind\ind \(h(x,n) = \sum_{i=0}^{-i-1} h_i(x,i). \)\\[.03in]
Clearly $h$ is recursive and nondecreasing and\\[.03in]
\ind\ind \(p(x0) + p(x1) \leq p(x) \leq \sum_{i=0}^{\oo}2^{-i-1} = 1 \)
                                           for all $x\in X^*$.\\*[.03in]
and hence $p$ is an incrementable predictor. Furthermore if $p'$
is any incrementable predictor then there is an $i\in N$
such that $p' = p_{i}$, so that $p(x)\geq 2^{-i-1}p'(x)$
for all $x \in X^*$. The theorem follows with $c = i+1$. $\Box$

\noind
{\bf Corollary} (Levin). For any optimal predictor $p$
and recursive distribution $p'$, the infinite sequences $z$
with absolutely bounded $rp(z(n))-rp'(z(n))$ are the sequences
which are M-L random relative to $p'$. In particular, the infinite
sequences whose redundancy relative to $p$ is absolutely
bounded are the M-L random sequences.

\noind
\UL{Proof.} $rp(z(n))-rp'(z(n))$ is bounded above for any 
sequence $z$ \In  $X^{\oo}$ which is M-L random relative to $p'$,
by Th.~5. Clearly it is also bounded below for if it
were not then $rp'(z(n))-rp(z(n))$ would not be bounded above,
in contradiction with Th.~8. $\Box$

 This section will be concluded with a proof of the fact
that any optimal universal predictor assigns nonvanishing
limit probabilities precisely to the recursive sequences.
This interesting result was previously obtained by
de Leeuw et al (1956) in a paper on probabilistic machines,
and was given its present interpretation by Levin (in
Zvonkin \& Levin, 1970). Levin's own proof depended on
properties of Loveland's ``uniform complexity" (Loveland, 1970).
For a closely related result see also Chaitin (1976).

\noind
{\bf Theorem 9} (de Leeuw et al., Levin). If $p$ is an optimal universal
predictor then $z\in  X^{\oo}$ is recursive iff $\limn p(z(n)) > 0$.

\noind
\UL{Proof.} \RA: For any recursive $z\in X^{\oo}$ define
$p'(z(n)) = 1$ for all $n\in N$ and $p'(x) = 0$ for $x \not\ppx z$.
By Th.~5 there exists a constant $c$ such that $rp'(z(n)) -
rp(z(n)) = -\log p(z(n)) \leq c$ for all $n\in N$, or $p(z(n))
\geq 2^{-c}$ for all $n\in  N$.\\*[.03in]
\LA: Let $z\in X^{\oo}$ be any sequence such that 
for some $c\in N$ $p(z(n)) > 2^{-c}$ for all $n\in N$.
Thus $\limn p(z(n))\geq 2^{-c}$. There are fewer than
$2^{c}$ such infinite sequences, i.e., with limiting 
probability $\geq 2^{-c}$. Let $z(k)$ be the shortest 
prefix of $z$ which is not a prefix of any of the other
sequences with limiting probability $\geq 2^{-c}$;
let $l$ be the smallest integer such that all finite sequences
of length $> l$ with probability $> 2^{-c}$ are prefixes of
infinite sequences with limiting probability $\geq 2^{-c}$
(it is not hard to see that such an $l$ exists -- note that if
there were infinitely many finite sequences whose
probabilities exceed $2^{-c}$ but which are not prefixes 
of a finite number of infinite sequences, then $p(\nil) = \oo$
would hold); finally, let $m = \max\{k,l\}$. Then $z(n)$ can
be computed for any $n > m$ by enumerating pairs $\lc x,q\rc
\in S$, where $S$ underlies $p$, until a pair is obtained 
such that $|x| = n$, $z(m)\ppx x$, and $q > 2^{-c}$; 
this $x$ is $z(n)$. $\Box$

 Note that another way of stating Th.~9 is that if
$p$ is any optimal universal predictor then $z\in X^{\oo}$ is
recursive iff there is a $c\in N$ such that $rp(z(n))
\geq n - c$ for all $n\in N$, i.e., the recursive 
sequences are those whose redundancy as a function
of initial-segment length $n$ is approximately $n$.

\subsection*{4. Predictability and Schnorr Randomness}
 Schnorr (1971) drew attention to the fact that 
the criteria of effectiveness employed by Martin-L\"{o}f
in defining sequential tests are too weak by the 
standards of constructive mathematics. In particular,
the measure $\mu V_{m}X^{\oo}$ is not in general a
recursive function of $m$ for a M-L sequential test $V$.
Putting the criticism another way, Schnorr pointed out
that M-L tests classify as nonrandom certain sequences
whose nonrandomness cannot be effectively observed,
in any reasonable sense of effective observation.

 He therefore proposed several alternative ways
of strengthening the criteria of effectiveness for
distinguishing between random and nonrandom sequences,
and showed that these lead to equivalent characterizations
of randomness (Schnorr, 1971, 1973). One proposal
is to require $\mu V_{m}X^{\oo}$ to be a recursive 
function of $m$. The random sequences are taken to be
those with bounded critical level relative to all
such tests. An equivalent proposal is contained in
the definition of Schnorr randomness which follows.

 A \UL{growth function} is a recursive nondecreasing
unbounded function $g:N\ra N$. A sequence $z\in X^{\oo}$
is \UL{Schnorr (S-) random} iff there does not exist 
a M-L sequential test $V$, a recursive lower bound $f$
of $m_{V}$, and a growth function $g$ such that
$\limsupn f(z(n))/g(n) > 0$.

 Evidently this definition expresses a specific thesis
about what it means for the nonrandomness in a
sequence to be effectively observable, viz., one must
be able to confirm effectively that the upward
excursions of the critical level are bounded below
by some growth function. (Note incidentally that one
could equally well use 1 or any other positive constant
in place of 0 on the right-hand side of the 
inequality).\footnote
  {Schnorr's contention that the S-random sequences best
  correspond to the intuitively random sequences is not
  universally accepted. Indeed M\"{u}ller (1972)
  proposes {\it weaker} criteria of effectiveness
  than Martin-L\"{o}f, citing the existence of M-L random
  sequences which are limiting recursively computable as
  a defect of Martin-L\"{o}f's conception of randomness.}

 It has already been noted that there is no universal
recursive predictor, and hence certainly no optimal
universal recursive predictor. Nevertheless the
concept of optimality, i.e., maximization of initial-segment
redundancy, is of as much interest in connection with recursive
predictors as in connection with incrementable predictors.

 In Sec.~3 an optimal predictor was required to
maximize initial-segment redundancy apart from a
constant. Adherence to Schnorr's criteria of effectiveness
calls for a slightly weaker notion of optimality. 
A recursive distribution $p$ is said to be
\UL{weakly optimal for $z$} where $z\in X^{\oo}$, iff
for every recursive distribution $p'$ and  every growth
function $g$, $\limsupn (rp'(z(n)) - rp(z(n)))/g(n) \leq 0$.
Thus no recursive distribution reveals the redundancy
of $z$ noticably better than a predictor which is weakly
optimal for $z$.

 For example, corresponding to any computable $r\in R$,
the predictor $p(x) = r^{n(x)}(1-r)^{|x|-n(x)}$,
where $n(x)$ = the number of ones in $x$, is weakly optimal
for all Bernoulli sequences with success probability
$r$, when the following notion of S-randomness relative
to a recursive distribution is substituted in Martin-L\"{o}f's
definition of Bernoulli sequences.

 A sequence $z\in X^{\oo}$
to be \UL{S-random relative to $p$} where $p$ is a
recursive distribution, iff there is no M-L sequential
test $V$ for $p$, recursive lower bound $f$ of $m_{V}$ and
growth function $g$ such that $\limsupn f(z(n))/g(n) > 0$.

 Once again a direct connection between optimality
and randomness can be established, much as in Th.~5.

 First it should be noted that the numerosity condition
of a M-L test $V$ for $p$ can be restated as\\*[.03in]
\ind\ind \(\sum_{y\in Y\cap V_m} \leq 2^{-m} \)\\*[.03in]
for every finite pf $Y \pss X^*$. Also the following
fact will be used.

\noind
{\bf Lemma 2.} Let $V$ be a M-L sequential test
for a distribution $p$, and let $g$ be a growth function.
Then for any $x \in X^*$, any integer $k\geq |x|$, and any
finite pf $Y\pss xX^{k-|x|}X^*$,\\*[.03in]
\ind\ind \( \sum_{y\in Y\cap V_{g(|y|)}}
           g(|y|)p(y) \ \leq \ (g(k)+1)2^{-g(k)} . \)

\noind
\UL{Proof.} By the numerosity condition the total
probability of sequences in $Y\cap V_{g(k)}$ is at
most $2^{-g(k)}$, that of sequences in $Y\cap V_{g(k)+1}$
at most $2^{-g(k)-1}$, etc. Hence the above sum is 
at most\\*[.03in]
\ind\ind $g(k)2^{-g(k)-1} + (g(k)+1)2^{-g(k)-2} + ... =
(g(k)+1)2^{-g(k)}$. $\Box$

\noind
{\bf Theorem 10.} For any recursive distribution $p$
and any $z\in X^{\oo}$, $z$ is S-random relative to $p$ iff
$p$ is weakly optimal for $z$.

\noind
\UL{Proof.} The proof parallels that of Th.~5.\\
\RA: Suppose that $p$ is not weakly optimal for $z$, i.e., 
there is a recursive distribution $p'$ and a growth
function $g$ such that $\limsupn[rp'(z(n))
- rp(z(n))]/g(n) > 0$. The test $V$ is defined as in Th.~5,
and its required properties established as before.
Thus $\ex c> 0$: for infinitely many $n$: 
$\log p'(z(n)) - \log p(z(n)) > cg(n)$, and hence for
infinitely many $n$: $m_{V}(z(n))/g(n) > 0$.
Since $m_{V}$ is recursive, this implies that $z$ 
is not S-random relative to $p$.\\*[.03in]
\LA: Suppose that $V$ is a M-L test for $p$ such that 
for some recursive lower bound $f$ of $m_{V}$ and some
growth function $g$, $\limsupn f(z(n))/g(n) > 1$.

 From this point on the situation is more complicated
than in Th.~5, because $\limn q(x,n)$ need not be recursive, 
even with $f$ replacing $m_{V}$ in the definition of $q$.
The limit operation must somehow be cut short, without
violating (sub-) additivity.

 Let $p'(x) = p_{1}(x) + p_{2}(x) + p_{3}(x)$
for all $x\in X^*$, where $p_{1}$, $p_{2}$, and
$p_{3}$ are defined recursively as follows:\\*[.03in]
\ind\ind $p_1(\nil) = 0$, \ $p_2(\nil) = (g(1)+1)2^{-g(1)}$,
                        \ $p_3(\nil) = 1 - p_2(\nil)$,\\
and for all $u\in X$ and $x\in X^*$\\
\ind\ind \( p_1(xu) = \max_{{\rm pf} \ Y\sseq\,xu(X^*-X^{k-|x|-1}X^*)}
                    \sum_{\stackrel{y\in Y}{f(y)\geq g(|y|)}} g(|y|)p(y), \)\\
\ind\ind \( p_2(xu) = (g(k)+1)2^{-g(k)}, \) \ and\\
\ind\ind \( p_3(xu) = p_3(x)/2 - (g(k)+1)2^{-g(k)}
                               + [p_1(x)+p_2(x)-p_1(x0)-p_1(x1)]/2, \)\\[.03in]
\noind
where $k$ = least integer $\geq |x| + 2$ such that 
$(g(k)+1)2^{-g(k)} < p_{3}(x)/2$.
$p'$ will be shown to be a recursive distribution, 
assuming without loss of generality that $g(n)\geq 2$
for all $n\in N$.

 The properties $p'(\nil) = 1$, additivity, and 
recursiveness are easily verified assuming that 
a suitable $k$ exists for each $x \in X^*$.
To prove the correctness of the latter assumption
by induction, assume that for all $x$ of length $l$ or less\\*[.03in]
\ind (a) $p_{1}(x), \ p_{2}(x)$, and $p_{3}(x)$ are well-defined; and\\
\ind (b) $p_{3}(x) > 0$.\\*[.03in]
Consider any particular $x$ of length $l$.
Denote the corresponding value of $k$ by $k'$ (if $x=\nil$, use $k'=1$).
Since $g$ is a growth function,
$(g(k)+1)2^{-g(k)} \ra 0$ as $k \ra \oo$. Also $p_{3}(x) > 0$,
hence the new value of $k$ required in the definition of 
$p'(xu), u\in\,X$, exists. Denote it by $k''$. Thus
$p_{1}(xu)$, $p_{2}(xu)$, and $p_{3}(xu)$ are well-defined.
From the definition of $p_{3}$, observe
that $p_{3}(xu)>0$ if $p_{1}(x) + p_{2}(x)%
\,\geq \ p_{1}(x0) + p_{1}(x1)$. But 
\begin{eqnarray*} 
 \lefteqn{p_{1}(x0) + p_{1}(x1) }\\
 & = & \max_{{\rm pf} \ Y \sseq\;xX(X^* - X^{k'' - |x - 1}X^*)}
            \sum_{\stackrel{y\in Y}{f(y)\geq g(|y|)}} g(|y|)p(y) \\
 & \leq & \max_{{\rm finite \ pf} \ \sseq\;xX^*}
            \sum_{\stackrel{y\in Y}{f(y)\geq g(|y|)}} g(|y|)p(y) \\
 & \leq & \max_{{\rm pf} \ Y \sseq\;x(X^* - X^{k'-|x|}X^*)}
            \sum_{\stackrel{y\in Y}{f(y)\geq g(|y|)}} g(|y|)p(y) \\
        & & + \max_{{\rm finite \ pf} \ \sseq\;xX^{k'-|x|}X^*}
            \sum_{\stackrel{y\in Y}{f(y)\geq g(|y|)}} g(|y|)p(y) \\
 & \leq & p_1(x) + (g(k')+1)2^{-g(k')} \ \ {\rm by \ Lemma \ 2} \\
 & = & p_1(x) + p_2(x),
\end{eqnarray*}

\noind
so that the inductive step is complete.
The induction starts at $l = 0$. In this case
(a) certainly holds and $p_{3}(\nil) 
\geq 1/4$, since $g(1)\geq 2$ by assumption.

 Now the definition of $p_{1}$ ensures that $p'(x) 
\geq g(|x|)p(x)$ whenever $f(x)\geq g(|x|)$. But $f(z(n))
> g(n)$ for infinitely many $n$, hence $p'(z(n))\geq
g(n)p(z(n))$ for infinitely many $n$, i.e., $\limsupn
[rp'(z(n)) - rp(z(n))]/log~g(n) \geq
1$, so that $p$ is not weakly optimal for $z$. $\Box$

 In conformity with Schnorr's notion of effectively
observable growth, a sequence $z\in X^{\oo}$ is defined
to be \UL{predictable} iff there exists a recursive
predictor $p$ and a growth function $g$ such that $\limsupn
rp(z(n))/g(n) > 0$; i.e., the redundancy of $z$
grows ``noticably". Note that $p$ may as well be taken
to be a recursive distribution, by Th.~1(b).
Also, since $(p(x)+2^{-|x|})/2$ defines a positive
recursive distribution, $p(x)$ may be taken to be 
positive for all $x\in X^*$. This leads to the following

\noind
{\bf Corollary.} A sequence $z\in X^{\oo}$ is S-random iff
it is not predictable.

\noind
\UL{Proof.} Let $p$ in Th.~10 be the uniform distribution. $\Box$

 An alternative proof of the corollary is easily obtained
from one of Schnorr's (1971) results about martingales.
Such a proof is given below, partly because of the
significance of the corollary and partly because of
the inherent interest of Schnorr's result.
Martingales (first used by Ville, 1939)
describe the capital of a gambler who bets on the
occurrence of 0 and 1 as next digit in a sequence and
subsequently wins the amount wagered on the digit
which actually follows and loses the amount wagered
on its complement. Formally, a \UL{martingale} is a
total function $f:\,X^*\ra \Rr^{+}$ such that $f(x) = (f(x0)
+ f(x1))/2$ for all $x\in X^*$. Schnorr showed that a
sequence $z\in X^{\oo}$ is S-random iff there does not
exist a recursive martingale $f$ and a growth function $g$
such that $\limsupn f(z(n))/g(n) > 0$. Again this
embodies the previous notion of effectively observable
growth, in this case of the gambler's capital.

\noind
\UL{Alternative proof of corollary.} \RA: Suppose that there
is a positive recursive distribution $p$ and a growth
function $g$ such that $\limsupn rp(z(n))/g(n) > 0$.
Then $f$ defined by $f(x) = 2^{|x|}p(x)$ for all
$x\in X^*$ is a recursive martingale and $\limsupn
f(z(n))/2^{g(n)} = \limsupn 2^{rp(z(n))}/2^{g(n)}
> 0$. Hence $z$ is not S-random.\\*[.03in]
\LA: Suppose that there is a recursive martingale $f$
and a growth function $g$ such that $\limsupn f(z(n))/g(n)
> 1$. Then $p$ defined by $p(x) = 2^{-|x|}f(x)/f(\nil)$
for all $x\in X^*$ is a recursive distribution and 
$\limsupn rp(z(n))/\log g(n) = \limsupn (\log~f(x)
- \log~f(\nil))/\log g(n) > 0$. Hence $z$ is predictable. $\Box$

 It is tempting, in view of Ths.~5 \& 10, to identify
optimal methods of prediction with ``rational" methods
of prediction, i.e., to stipulate as a general requirement
for any ``rational" method of probabilistic prediction that
any sequence of observations to which the method is applied
should appear to behave randomly relative to the probability
assignments of the method. This seems to be Levin's view
(Zvonkin \& Levin, 1970, and Levin, 1973, 1976), since he identifies
the probabilities of the optimal semicomputable measure with
intuitive prior probabilities. However, this requirement is
surely too strong, since no universal (and hence no generally
optimal or weakly optimal) computable predictor exists; i.e.,
no predictive method exists which is both effective and 
``rational" in so strong a sense.

 Therefore it seems appropriate to admit as ``rational"
those methods of prediction relative to which any sequence
passes some, but not necessarily all, tests for randomness. 
In particular, it may be sufficient to require all sequences
to satisfy a law of large numbers relative to the predictive
method. Then it becomes possible to construct effective
methods of prediction which are ``rational".

 Th.~10 is thus best viewed as providing a method for
comparing ``rational" predictors as to their predictive
``power": a predictor is powerful to the extent that
sequences pass a variety of randomness tests (in addition
to those criterial to ``rationality") relative to the 
pedictor. The second part of the proof of Th.~10 also
indicates how a predictive method might be improved if
a sequence is known which does not pass some randomness
test relative to the method.

 What kind of law of large numbers can be formulated
relative to an arbitrary recursive predictor? The law
should express that the frequencies of the predicted 
events conform with the probabilities assigned to them.
Specifically, the frequency of occurrence of events
which are assigned conditional probabilities within some
particular interval should lie in that same interval.
The next theorem shows that this law is satisfied by
relatively S-random sequences.

 First a \UL{rational-computable} distribution is
defined as one whose values are rational numbers 
which can be found uniformly effectively for all
arguments. Note that if $p$ is a recursive distribution
then there is a rational-computable approximation $p'$ to $p$
which attributes the same redundancy as $p$ to all $x\in X^*$,
apart from a constant.
$p'(x)$ need only lie within distance $2^{-|x|}p(x)$ of $p(x)$
for all $x\in X^*$; this guarantees that $\ln p(x) - \ln p'(x) < 2$
for all $x \in X^*$. Hence if $p$ is weakly optimal for a
sequence, so is $p'$. Thus the restriction to rational-%
computable distributions in the following theorem
detracts little from its interest. It is assumed that an
assertion in square brackets has numerical value 1 if the 
assertion is true and value 0 otherwise.

\noind
{\bf Theorem 11.} Let $p$ be a rational-computable
distribution which is weakly optimal for some $z\in X^{\oo}$.
Let $r$, $s$ be rational numbers such that $.5 < r\leq s < 1$ 
and\\*[.03in]
\ind\ind \( \liminfn \sum_{i=0}^n \sum_{w\in X} 
                     [r\leq p(z(i))w)/p(z(i)) \leq s]/g(n) > 0 \)\\*[.03in]
for some growth function $g$, i.e., the frequency
of next-digit predictions with probabilities in
$[r,s]$ is nonnegligible. Then the proportion
of {\it confirmed} predictions with probabilities 
in this range satisfies

\[ {\rm lim} \{\begin{array}{c}{\rm inf}_n \\{\rm sup}_n \end{array}\}
           \frac{\sum_{i=1}^n [r\leq p(z(i+1))/p(z(i))\,\leq s]}
           {\sum_{i=0}^n \sum_{y\in X}[r\leq p(z(i)y)/p(z(i))\,\leq s]}
              \{\begin{array}{c} \geq r \\ \leq s \end{array}\}.\]
\noind
\UL{Proof.} In view of Th.~10 one might attempt a proof
by formulating a suitable M-L test for $p$. Instead the following
argument proceeds directly from the assumption that $p$ is 
weakly optimal for $z$.

 Suppose contrary to the theorem that the lim inf
of the above ratio $< r - a$ for some real $a > 0$.
As before let $x^-$ denote the sequence obtained
by changing the last digit of $x\in XX^*$ to its complement.
Let $b$ be any rational number such that $0 < b < a$,
and define $p'$ by\\*[.03in]
\ind\ind $p'(\nil) = 1$ and for all $x\in X^*,\,u\in X$\\*[.03in]
\ind\ind $p'(xu)/p'(x) = p(xu)/p(x) - b$ if $p(xu)/p(x)\;\in [r,s]$,\\
\ind\ind $p'(xu)/p'(x) = p(xu)/p(x) + b$ if $p(xu^-)/p(x)\;\in [r,s]$,\\
\ind\ind $p'(xu)/p'(x) = p(xu)/p(x)$ otherwise,\\*[.03in] 
where any occurrences of 0/0 are replaced by 0.
Clearly $p'$ is a recursive distribution (note that
$p(xu)/p(x)\in [r,s]$ is decidable if $p$ is rational-computable).

 By supposition, for infinitely many $n\in N:\,\ex k,\,l\in N:$\\*[.05in]
\ind\ind \(k = \sum_{i=0}^{n-1} [r\leq p(z(i+1))/p(z(i))\leq s],\)\\*[.05in]
\ind\ind \(l = \sum_{i=0}^{n-1} [r\leq p(z(i+1)!13)/p(z(i))\leq s]\;> 0,\)
                                                                and\\*[.03in]
\ind\ind $k/(k+l) < r - a$.\\*[.03in]
For any such $n$ denote the $k$ values of $p(z(i+1))/p(z(i))$
in $[r,s]$ by $p_{1},...,p_{k}$ and the $l$ values of
$p(z(i+1))/p(z(i))$ in $[1-s,1-r]$ by $q_{1},...,q_{l}$
(observe that $p(z(i+1))/p(z(i))\in [1-s,1-r]$ iff
$p(z(i+1)!13)/p(z(i))\in [r,s]$).
The corresponding values of $p'(z(i+1))/p'(z(i))$ are
$p_{1}-b,...,p_{k}-b$ and $q_{1}+b,...,q_{l}+b$
respectively. Since the conditional probabilities of 
the digits of $z$ determined by $p$ and $p'$ for $i\leq n$
are otherwise identical and nonzero,
\begin{eqnarray*}
\lefteqn{\ln p'(z(n) - \ln p(z(n))}\\
  & > & \sum_{i=1}^k \frac{b}{p_i-b}\;+\;\sum_{i=1}^l \frac{b}{q_i+b} \\
  & \geq & -\frac{kb}{r-b} + \frac{lb}{1-r+b} \ 
              = \ b\frac{(k+l)(r-b)-k}{(r-b)(1-r+b)}\\
  & > & b\frac{(k+1)(a-b)}{(r-b)(1-r+b)} \ {\rm since} \ (k+1)r - k > (k+1)a.
\end{eqnarray*}
\noind
But $k+l$, as a function of $n$, is bounded below
by $g(n)$ times some constant for all sufficiently
large $n$, hence\\*[.03in]
\ind\ind $\limsupn [rp'(z(n)) - rp(z(n))]/g(n) > 0$,\\*[.03in]
in contradiction with the assumption that $p$ is weakly optimal for $z$.
The proof of the second part of the theorem is
entirely analogous to that of the first. $\Box$

 It may be possible to strengthen the theorem in various ways,
for example by specifying the rate of convergence
towards the two limits (more precisely, by specifying
the critical levels associated with deviations of
the proportion of confirmed predictions from lower
bound $r$ or upper bound $s$). It may also be possible to formulate
the theorem so as to apply to arbitrary subsequences 
extracted from infinite sequences by application of
recursive selection rules, or to apply to prediction
of more general types of events, such as arbitrary
cylinder sets of sequences. However, the theorem
in its present form sufficiently illustrates the
conformity between the predictions of an optimal
predictor and the occurrences of the predicted events.

\subsection*{5. Solomonoff Predictors}
 In this section incrementable predictors will be related
to one of the classes of sequence predictors proposed
by Solomonoff (1964) in his pioneering work on inductive
inference for infinite sequences (see also Willis, 1970,
Zvonkin \& Levin, 1970, Chaitin, 1975, Cover, 1974, 
Leung-Yan-Cheong \& Cover, 1975, and Solomonoff, 1976
for closely related studies).

 Actually Solomonoff considered four methods of
predicting sequences probabilistically.
In the first three methods the probability $p(y)$ of
a sequence $y\in X^*$ is obtained by summing terms of 
the type $2^{-|x|}$, where $x$ is an encoding or
program from which $y$ can be generated on a fixed
machine. This formalizes the intuitive idea that
the highest prior probability should go to sequences
with short and/or numerous encodings. The three
methods differ in the types of machines considered
and in other relatively minor respects. Solomonoff
conjectures that they are essentially equivalent.

 Here a machine-independent formulation of Solomonoff's
second method will be used. The formulation is
based on Schnorr's notion of a \UL{process} 
(or \UL{monotone function} defined as a partial
recursive function $f:\,X^*\ra X^*$ such that $f(x)\px f(xy)$
for all $x,\; xy$ in the domain of $f$ (Schnorr, 1971, 1973; also
Zvonkin \& Levin, 1970).
Thus processes map extensions of inputs into extensions
of outputs. $x$ is said to be an \UL{encoding} of $y$ relative
to process $f$, abbreviated as $f(x)\codes\,y$, whenever $f$
maps $x$, but no proper prefix of $x$, into an extension of $y$. 
In symbols, $f(x)\codes\,y$ iff $x\in f^{-1}(yX^*) - f^{-1}(yX^*)XX^*$.
A \UL{Solomonoff predictor} is now defined as a function $p:\,X^*\ra R$
such that for some process $f$
\[ p(y) = \sum_{f(x)\codes\,y} 2^{-|x|} \ {\rm for \ all} \ y\in X^*, \]
\noind
where a sum over no terms is 0, as before.
The symbol $p_{f}$ denotes the 
Solomonoff predictor determined by any process $f$.
The notation $\sig S$ will be used as an abbreviation for
 \( \sum_{x\in S} 2^{-|x|}. \)

Thus $p_{f}(y) = \sig\{x|f(x)\codes\,y\}$.
Note that for any pf set $S\pss X^*,\, \mu SX^{\oo} = \sig S$,
where \Mu\ is the uniform measure on $X^{\oo}$.

 The equivalence of Solomonoff predictors and incrementable
predictors will now be established. Thus in considering
methods of probabilistic sequence prediction, one can in
principle restrict attention to methods which attribute
high probability to sequences with short and/or numerous
encodings, just as Solomonoff suggested.

 In the following it will sometimes be helpful
to think of any set S \Sseq X* as a set of nodes of
a binary tree rooted at \Nil, with a pf set containing
leaf nodes only. At other times it will be useful to think of a
sequence $x = x_{1}x_{2}...x_{n}$ ($x_{i}\in X$)
as the real {\it interval}\\*[.05in]
\ind\ind \( [\,\sum_{i=1}^n x_i 2^{-i}\;,\;\sum_{i=1}^n x_i 2^{-i} 
                                                     + 2^{-n} ), \)\\*[.03in]
i.e., the least interval containing all numbers in $B$
whose (finite) radix-2 representation begins with $x$. Then pf
sets of sequences correspond to sets of disjoint intervals, 
and extensions of sequences correspond to subintervals.

\noind
{\bf Theorem 12} (Levin). A function $p:X^*\ra R$ is an
incrementable predictor iff it is a Solomonoff predictor.

\noind
\UL{Proof.} \LA: For any $y\in X^*$, sequences mapped by
a process $f$ into extensions of $y0$ or $y1$ are certainly
mapped into extensions of $y$. Hence $p_{f}(y0) +
p_{f}(y1)\leq p_{f}(y)\leq 1$. $p_{f}$ is 
incrementable because $f$ is re and for any finite process
$f'\sseq f$, $p_{f'}(y)$ is rational and $\leq p_{f}(y)$
for all $y\in X^*$.\\*[.03in]
\RA: As previously indicated$^2$, the ranges of the 
functions underlying incrementable predictors might
equally well have been confined to $B$ (instead of $Q$).
Also, by Lemma 1 the function $h$ underlying an
incrementable predictor can be chosen to be subadditive,
i.e., $h(y,n)\geq h(y0,n)+h(y1,n)$ for all $y\in X^*$, $n\in N$.
The elements of a process $f$ such that $p_{f} = p$ can
now be generated as follows. At stage $n$ of the generation
procedure a finite set of elements is added to the process
for each $y$ of length 0,~1,~...,~$n$ (in that order), so as
to increase the Solomonoff probability of $y$ from $h(y,n-1)$
to $h(y,n)$. Each new element $\lc x,y\rc$ added to the process
is chosen so that no extension of $x$ is as yet in the 
domain of the process, and $x$ properly extends some $x'$ 
where $\lc x',y(|y|-1)\rc$ was added to the process earlier
(the latter condition is omitted for $y = \nil$).
The required sets of additions to increase the Solomonoff
probabilities of $y0$ and $y1$ from $h(y0,n-1)$ and $h(y1,n-1)$
to $h(y0,n)$ and $h(y1,n)$ respectively always exist because
of the subadditivity of $h$ (a detailed argument using 
induction on $n$ and $|y|$ is easily supplied). The sets are
always finite because the required probability increments
are in $B$. The construction is illustrated in Fig.~1,
using the interval representation of certain sequences in
the domain of the process being constructed. $\Box$

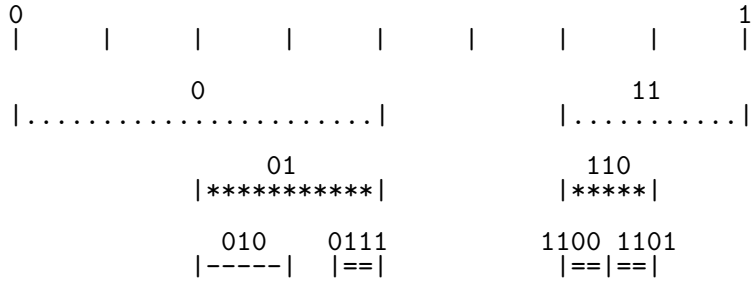
\renewcommand{\baselinestretch}{0.7} 

\begin{verbatim}
       0                                               1
       |     |     |     |     |     |     |     |     |

                   0                            11
       |.......................|           |...........|

                        01                   110
                   |***********|           |*****|

                     010    0111          1100 1101
                   |-----|  |==|           |==|==|
\end{verbatim}

\renewcommand{\baselinestretch}{0.8} 

 \hspace*{5ex}Graphical symbolism:\\*[.03in]
 \hspace*{10ex}\verb+|.....|+ mapped into proper prefixes of $y$\\
 \hspace*{10ex}\verb+|*****|+ mapped into $y$\\
 \hspace*{10ex}\verb+|-----|+ mapped into $y0$\\
 \hspace*{10ex}\verb+|=====|+ mapped into $y1$
\begin{itemize}
\item[Fig.\,1.] Construction of process $f$ corresponding to given
             incrementable predictor. At the point shown
             $p_{f}(y)=3/8,\,p_{f}(y0)=1/8, \,p_{f}(y1)=3/16$. 
\end{itemize}

\noind
A process $f$ is said to be \UL{universal} iff
for every process $f':\,\ex x\in X^*:\,\lm y\,f(xy) = f'$. Levin
(in Zvonkin \& Levin, 1970) and also Schnorr (1973) proved the existence 
of a universal process.

\noind
\UL{Corollary.} If $f$ is a universal process then $p_{f}$
is an optimal universal predictor.

\noind
\UL{Proof.} Let $p$ be an optimal universal predictor
and $f'$ a process such that $p_{f'} = p$. Then 
$\ex x\in X^*:\,\lm y\,f(xy) = f'$. Hence\\*[.03in]
\ind\ind \( p_{f'}(z) = \sig\{y|\,f(xy)\codes\,z\} = 
         2^{|x|}\sig\{xy|f(xy)\codes\,z\} \leq 2^{|x|}p_f(z) \)\\*[.03in]
for all $z\in X^*$, and the corollary follows.$\Box$

 An optimal Solomonoff predictor, as defined in the 
corollary, is not quite the same as Cover's universal
prediction scheme (Cover, 1974). Cover's scheme could
be obtained by retaining only certain terms of $p_{f}$,
namely those contributed by encodings $x$ such that $x$ is
the {\it shortest} argument for which $f$ assumes value $f(x)$. 
Optimal Solomonoff predictors do appear to be essentially
the same as Solomonoff's own measures $P'_{M}$ (Solomonoff,
1976) apart from the normalization term employed by
Solomonoff. To prove this one would have to relate processes
to Solomonoff's computational model, which permits finite
and infinite inputs and outputs, as well as finite outputs 
generated by nonterminating computations.

 Willis (1970) called a distribution $p$ 
\UL{binary-computable} iff it is a recursive mapping into $B$.
As in the case of rational-computable distributions
it is easily seen that any recursive distribution $p$ can
be approximated by a binary-computable distribution which
attributes the same redundancy as $p$ to all $x\in X^*$, 
apart from an arbitrarily small constant.

 A class of processes will now be characterized which
corresponds to the class of binary-computable distributions.
This leads to a machine-independent formulation of 
one of Willis' main results (a closely related result is proved 
in Zvonkin \& Levin, 1970).

 A process $f$ is called \UL{endless} iff the set
$f({z(n)|n\in N})$ is infinite for every $z\in X^{\oo}$.

\noind
{\bf Theorem 13.} $p$ is a binary-computable distribution
iff there is an endless process $f$ such that $p = p_{f}$.

\noind
\UL{Proof.} \RA: A procedure for generating a process 
can be used, similar to that in the proof of Th.~12.
Corresponding to each $y$ of length 0,~1,~2,~... (considered
in that order), a set of elements is added to the process 
such that the Solomonoff probability of $y$ becomes $p(y)$.
Because of the distribution property, i.e., $p(y) = 
p(y0)+p(y1)$, the construction of $f$ and the proof that
$p_{f} = p$ present no difficulty. Now clearly the minimal
length of sequences in the successive subdomains $f^{-1}({\nil}),
f^{-1}(X), ...~, f^{-1}(X^{n})$, ... is strictly increasing
as a function of $n$, and $X^{\oo} = f^{-1}({\nil})X^{\oo} =
f^{-1}(X)X^{\oo} = ...~$. Hence $f$ is endless.\\*[.03in]
\LA: For a given endless process $f$, $p(y)$ can be computed
for any $y\in X^*$ by enumerating elements of $f$ until a finite
subprocess $f'\sseq f$ is obtained such that\\*[.05in]
\ind\ind \( \sum_{|x|=|y|} p_{f'}(x) = \sum_{|x|=|y|} p_f(x) = 1. \)\\*[.05in]
At that point $p_{f}(y)$ will be available as a finite
sum of nonpositive powers of 2. To see that the required
$f'$ always exists, note first that $f^{-1}({y})$ is finite
for all $y$. For if it were not for some $y$, then by K\"{o}nig's
infinity lemma (see e.g., Knuth, 1968) there would be an
infinite sequence $z$ all of whose prefixes have extensions
in $f^{-1}({y})$. Hence by the definition of an endless 
process there would be an infinite set of prefixes of $z$
mapped by $f$ onto an infinite set of output sequences.
But prefixes of $z$ can only be mapped into the finite set of
prefixes of $y$ (by the process property); thus $f^{-1}(\{y\})$
cannot be infinite. It follows that for any sufficiently
large $n$, no element of $X^{n}X^*\cap~{\rm dom}\,f$ will be mapped into
sequences of length $\leq |y|$. But since dom~$f$ contains 
arbitrarily long prefixes of every infinite sequence, hence
$(X^nX^*\cap~{\rm dom}~f)X^{\oo}$ = $X^{\oo}$ for all $n\in N$.
Thus the Solomonoff probability of the sequences generated
by $f$ on this subdomain is 1, and for $n$ sufficiently large
these sequences are all of length $\geq |y|$. Furthermore,
the set $X^{n}X^*\cap~{\rm dom}~f$, made pf by removal of sequences
which properly extend other sequences in the set, is finite;
otherwise there would be a $z\in X^{\oo}$ none of whose 
prefixes are in the set, again by K\"{o}nig's lemma.
Thus the subprocess $f'\sseq f$ with pf subdomain 
$X^{n}X^*\cap~{\rm dom}~f$ and with $n$ sufficiently large possesses
the required properties. $\Box$

\noind
{\bf Corollary 1.} $p$ is a binary-computable distribution
iff there exists an endless recursive process $f$ such that
$p = p_{f}$.

\noind
\UL{Proof.} It need only be shown that for every endless 
process $f$ there exists an endless recursive process $f'$
such that $p_{f'} = p_{f}$. Such an $f'$ is easily 
obtained by a slight modification of the process construction
mentioned in the first part of the proof of Th.~13 (the
construction is applicable because $p_{f}$ is a binary-%
computable distribution). In addition to the process
elements generated in that construction, $\lc x,y\rc$ is added to
the process whenever $\lc x0,yu\rc$ and $\lc x1,yv\rc$ have previously
been added, where $\{u,v\}\sseq X$. These additions do not
affect the Solomonoff probabilities, and are easily seen to
extend the domain of the process to $X^*$. $\Box$

 Willis also showed that if $p$ is a binary computable
distribution then there is a machine (of the type he
considered) whose {\it shortest} encoding for any output sequence
$y$ determines the highest-order digit of $p(y)$. Relative
to certain machines, therefore, sequence prediction on
the basis of the shortest encoding is nearly as accurate
as prediction on the basis of all encodings.\footnote
  {Solomonoff, Willis, and Chaitin have all commented on the
  relationship between sequence prediction and ``scientific"
  prediction. Willis' result about the efficacy of the shortest
  encoding seems related to the efficacy of the simplest
  (shortest) theory in scientific prediction.}

 An analogous but somewhat stronger result can be proved
to the effect that a process exists corresponding
to $p$ in which {\it each} digit of $p(y)$ is determined by 
exactly one encoding of $y$. To do so, however, the notion of
encoding used so far needs to be modified, as Willis'
result would be patently false on the basis of that
notion. For consider the predictor $p(1^{n}) = 1$
for all $n\in N$, all other values being zero. 
Although a process $f$ can be constructed such that
$\min\{x|\,f(x)\codes\,1^{n}\}$ grows arbitrarily slowly
with $n$, this minimum must nevertheless grow unboundedly
and hence the fractional contribution of any minimal
encoding of $1^{n}$ to $p_{f}(1^{n})$ must 
approach 0 as $n\ra \oo$.

 This difficulty in reformulating Willis' result
disappears if the following ``liberalized" notion
of encoding is used. $x$ is said to be a \UL{reduced encoding}
of $y$ (symbolically, $f(x)\rcodes y$), iff there is a finite pf
$S\pss X^*$ such that $\sig S = 1$ and $f(xS)\pss yX^*$, and no such
$S$ exists for any proper prefix of $x$.\footnote
  {A correspondence can be established between processes
  and Willis' concrete model of computation (Willis, 1970)
  such that ``$f(x)\rcodes\,y$" becomes equivalent to ``$x$ is an 
  $|x|$-program for $y$".}
This is still a reasonable notion of encoding, since it
is possible to generate $y$ given $|y|$ and a reduced
encoding of $y$.
Furthermore, encodings could be replaced by reduced 
encodings in the definition of Solomonoff probabilities, 
i.e., \\*[.05in]
\ind\ind \( \sig\{x|\,f(x)\rcodes\,y\} = \sig\{x|\,f(x)\codes\,y\} 
                                                           = p_{f}(y).\)

 In the following corollary ``$\ex_{1}$" denotes ``there
exists exactly one".

\noind
{\bf Corollary 2.} $p$ is a binary computable distribution iff 
there exists an endless recursive process $f$ such that
for all $y\in X^*$\\*[.05in]
\ind\ind \( p(y) = p_f(y) = \sum_{n=0}^{\oo}[\ex_1 x: |x|=n \ 
                                          \& \ f(x)\rcodes y]2^{-n}.\)

\noind
\UL{Proof.} Again only a slight modification of the
construction in the first part of Th.~13 is needed.
The modification ensures that the ``intervals" chosen
for $f^{-1}(\{y0,y1\})$ (see Fig.~1) finitely partition
the ``intervals" previously chosen for $f^{-1}(\{y\})$
in such a way that to each digit of $p(y0)$ or $p(y1)$
contributing $2^{-i}$ to the probability, there
corresponds a set of adjacent intervals whose union
represents some $i$-sequence. That such a partitioning
always exists can be proved by induction on $|y|$. $\Box$

\subsection*{6. Concluding Remarks}
 It has been shown that the notion of a predictor provides 
a common basis for the study of randomness and the study of
probabilistic sequence prediction. The random sequences are
those which are irredundant with respect to all effective 
predictors, and all effective predictors are obtained by
assigning high probabilities to sequences with short and/or
numerous encodings with respect to some effective process.
It was also suggested that a minimal constraint on any
``rational" method of prediction is that all sequences obey 
a law of large numbers relative to it, while the requirement 
that all sequences should appear to behave randomly relative
to it is too strong.

 A new proof of the existence of an optimal incrementable
predictor was given. The fact that this predictor is not
computable detracts from its ``practical" interest. Perhaps
more interesting than the optimal predictor itself is its
method of construction. Since this is based on the recursive
enumerability of the class of predictors under consideration,
a similar construction is possible for more restricted
classes of predictors, e.g., the predictors derived from the
primitive recursive functions. Thus there will be predictors
which are optimal within ``practical" classes of functions,
whenever the weighted sum of predictors stays within the
class under consideration. It should not be hard to prove
(or ensure) that such predictors are also ``rational".

 An open question is whether a process can be found 
corresponding to any incrementable predictor $p$ such that
the highest-order digit of $p(x)$, $x\in X^*$, is determined
by the {\it shortest} reduced encoding of $x$ relative to
the process. An affirmative answer would give the analogue
of Th.~13, Cor.~2 for incrementable predictors.
Another set of questions concerns the classification of
randomness tests according to the growth in redundancy of
sequences rejected by such tests. Such a classification should
be easily obtainable from Schnorr's classification of 
randomness tests according to the growth of martingales
(Schnorr, 1971b). An entirely different set of questions
arises if the {\it difficulty} of predicting sequences
which are predictable to some degree is investigated.
Some of Schnorr's (1971b) work on complexity-based degrees
of randomness pertains to these questions.\\

\noindent
{\bf Acknowledgements}  
 The author is indebted to Dr.~Amram Meir of the Department
of Mathematics and to Dr.~Arthur Wouk of the Department of 
Computing Science of the University of Alberta for steering
him away from a faulty version of Th.~13. The research was
supported by the National Research Council of Canada under
Operating Grant A8818.

\newpage
\subsection*{References}
\parskip 0in
\noind
ABRAMSON, N.\ (1963), {\it Information Theory and Coding}, \hh
McGraw-Hill, New York, N.Y. 

\noind
BLUM, L., and BLUM, M.\ (1973), ``Inductive inference: \hh
a recursion theoretic approach", Memo.\ No.\ ERL-M386, Electronics 
Res.\ Lab., Univ.\ of Calif., Berkeley; also, ``Towards a mathematical
theory of inductive inference," {\it Inform.\ Contr.\ 28}
(1975), 125-155.

\noind
 CHAITIN, G.\,J.\ (1966), \hh ``On the length of programs for
computing finite binary sequences," {\it J.\ Ass.\ Comp.\ Mach.\,13},
547-569.

\noind
 CHAITIN, G.\,J.\ (1969), \hh ``On the length of programs for
computing finite binary sequences: statistical considerations,"
{\it J.\ Ass.\ Comp.\ Mach.\,16}, 145-159.

\noind
 CHAITIN, G.\ J.\ (1970), \hh ``On the difficulty of computations,"
{\it IEEE Trans.\ Inf.\ Theory IT-16}, 5-9.

\noind
 CHAITIN, G.\ J.\ (1974), \hh ``Information-theoretic computational
complexity," {\it IEEE Trans.\ Inf.\ Theory IT-20}, 10-15.

\noind
 CHAITIN, G.\ J.\ (1975), \hh ``A theory of program size formally 
identical to information theory," {\it J.\ Assoc.\ Comp.\ Mach.\ 22},
329-340.

\noind
 CHAITIN, G.\ J.\ (1976), \hh ``Information-theoretic 
characterizations of recursive infinite strings,"
{\it Theoret.\ Comput.\ Sci.\ 2}, 45-48.

\noind
 CHAITIN, G.\ J.\ (1977), \hh ``Algorithmic information theory,"
{\it IBM J.\ Res.\ Develop.\ 21}, 350-359.

\noind
 CHURCH, A.\ (1940), \hh ``On the concept of a random sequence,"
{\it Bull.\ Amer.\ Math.\ Soc.\ 46}, 130-135.

\noind
 COVER, T.\ M.\ (1974), \hh ``Universal gambling schemes and
the complexity measures of Kolmogorov and Chaitin," 
Statistics Dept.\ Rep.\ 12, Stanford Univ., Stanford, CA.

\noind
 DE LEEUW, K., MOORE, E.\,F., SHANNON, \hh C.\,E., and SHAPIRO, N.\ 
(1956), ``Computability by Probabilistic Machines," Automata
Studies, Princeton Univ., Princeton, N.\,J.\ 

\noind
 GOLD, E.\ M.\ (1967), \hh ``Language identification in the limit,"
{\it Inform.\ Contr.\ 10}, 447-474.

\noind
 KNUTH, D.\ E.\ (1968), \hh {\it The Art of Computer Programming,
Vol.~1: Fundamental Algorithms}, Addison-Wesley, Reading, Mass.,
p 381.

\noind
 KOLMOGOROV, A.\ N.\ (1965), \hh ``Three approaches to the 
quantitative definition of information," {\it Inf.\ Transmission 
1}, 3-11; also {\it Int.\ J.\ Comp.\ Math.\ 2} (1968), 
157-168.

\noind
LEUNG-YAN-CHEONG, S.\ K., and COVER, T.\ M.\ (1975), \hh
``Some inequalities between Shannon entropy and Kolmogorov,
Chaitin, and extension complexities," Tech.\ Report no.\ 16,
Statistics Dept., Stanford Univ., Stanford, CA.;
to appear in {\it IEEE IT}. 

\noind
LEVIN, L.\,A.\ (1973), \hh ``On the notion of a random sequence,"
{\it Soviet Math.\ Doklady 14}, 1413-1416.

\noind
LEVIN, L.\,A.\ (1976), \hh ``Uniform tests of randomness,"
{\it Soviet Math.\ Doklady 17}, 337-340.

\noind
LOVELAND, D.\,W.\ (1970), \hh ``A variant of the Kolmogorov
notion of complexity," {\it Inform.\ Contr.\ 15},
510-526.

\newpage
\noind
MARTIN-L\"{O}F, P.\ (1966), \hh ``The definition of random
sequences," {\it Inform.\ Contr.\ 9}, 602-619.

\noind
M\"{U}LLER, D.\ W.\ (1972), \hh ``Randomness and extrapolation,"
{\it Proc.\ 6th Berkeley Symp.\ on Math.\ Statistics and 
Probability}, June~21 - July~18, 1970, Le Cam, L.\,M., Neyman, J.,
and Scott, E.~L.\ (eds), Univ.\ Calif.\ Press, Berkeley and 
Los Angeles, CA., 1-31.

\noind
ROGERS, H.\ (1967), \hh {\it Theory of Recursive Functions and
Effective Computability,} McGraw-Hill, New York, N.\,Y.

\noind
SCHNORR, C.\,P.\ (1971a), \hh ``A uniform approach to the 
definition of randomness," {\it Math.\ Systems Theory 5},
9-28.

\noind
SCHNORR, C.\,P.\ (1971b), \hh ``Zuf\"{a}lligkeit und Wahrscheinlichkeit,"
{\it Lecture Notes in Mathematics, Vol.\ 218}, Springer, Berlin -
Heidelberg - New York.

\noind
SCHNORR, C.\,P.\ (1973), \hh ``Process complexity and effective 
random tests," {\it J.\ Computer and Systems Sciences 7},
376-388.

\noind
SOLOMONOFF, R.\,J.\ (1964), \hh ``A formal theory of inductive
inference," {\it Inform.\ Contr.\ 7}, 1-22, 224-254.

\noind
SOLOMONOFF, R.\,J.\ (1976), \hh ``Complexity based induction
systems: comparisons and convergence theorems," Report RR-329,
Rockford Research, Cambridge, Mass.

\noind
VILLE, J.\ (1939), \hh {\it Etude Critique de la Notion Collectif,}
Gauthier-Villars, Paris.

\noind
VON MISES, R.\ (1919), \hh ``Grundlagen der Wahrscheinlichkeitstheorie,"
{\it Math.\ Z.\ 5}, 52-99.

\noind
WALD, A.\ (1937), \hh ``Die Widerspruchsfreiheit des
Kollektivbegriffs in der Wahrscheinlichkeitsrechnung,"
{\it Ergebnisse eines math.\ Koll.\ 8}, 38-72.

\noind
WILLIS, D.\,G.\ (1970), \hh ``Computational complexity and
probability constructions," {\it J.\ Ass.\ Comp.\ Mach.\ 17},
241-259.

\noind
ZVONKIN, A.\,K., and LEVIN, L.\,A.\ (1970), \hh ``The 
complexity of finite objects and the development of the
concepts of information and randomness by means of the 
theory of algorithms," {\it Russian Math.\ Surveys 25},
83-124.

\end {document}